\PassOptionsToPackage{numbers,sort&compress}{natbib}
\documentclass[preprintnumbers,amsmath,amssymb,prd, notitlepage,nofootinbib,twocolumn, superscriptaddress]{revtex4}

\usepackage{amsfonts,amssymb,amsmath}
\usepackage{gensymb}
\usepackage{color}
\usepackage{graphicx}
\usepackage{multirow}
\usepackage[utf8]{inputenc}
\usepackage{aas_macros}
\usepackage{hyperref}

\usepackage[normalem]{ulem}

\newcommand{\greaterthanapprox}{\mathrel{\vcenter{
  \offinterlineskip\halign{\hfil$##$\cr
    >\cr\noalign{\kern2pt}\sim\cr\noalign{\kern-2pt}}}}}
    
    \newcommand{\lessthanapprox}{\mathrel{\vcenter{
  \offinterlineskip\halign{\hfil$##$\cr
    <\cr\noalign{\kern2pt}\sim\cr\noalign{\kern-2pt}}}}}
    
\newcommand{\lb}{\left(}
\newcommand{\rb}{\right)}

\newcommand{\be}{\begin{equation}}        
\newcommand{\ee}{\end{equation}}

\begin{document}

\title{Converting dark matter to dark radiation does not solve cosmological tensions}

\author{Fiona McCarthy}
\email{fmccarthy@flatironinstitute.org}

\affiliation{Center for Computational Astrophysics, Flatiron Institute, New York, NY, USA 10010}

\author{J.~Colin Hill}
\affiliation{Department of Physics, Columbia University, 538 West 120th Street, New York, NY, USA 10027}
\affiliation{Center for Computational Astrophysics, Flatiron Institute, New York, NY, USA 10010}

\date{\today}

\begin{abstract}

Tensions between cosmological parameters (in particular the local expansion rate $H_0$ and the amplitude of matter clustering $S_8$) inferred from low-redshift data and data from the cosmic microwave background (CMB) and large-scale structure (LSS) experiments have inspired many extensions to the standard cosmological model, $\Lambda$CDM. Models which simultaneously lessen both tensions are of particular interest. We consider one scenario with the potential for such a resolution, in which some fraction of the dark matter has converted into dark radiation since the release of the CMB. Such a scenario encompasses and generalizes the more standard ``decaying dark matter'' model, allowing additional flexibility in the rate and time at which the dark matter converts into dark radiation. In this paper, we constrain this scenario with a focus on exploring whether it can solve (or reduce) these tensions. We find that such a model is effectively ruled out by CMB data, in particular by the reduced peak-smearing due to CMB lensing on the power spectrum and the excess integrated Sachs--Wolfe (ISW) signal caused by the additional dark energy density required to preserve flatness after dark matter conversion into dark radiation.  Thus, such a model does not have the power to reduce these tensions without further modifications.  This conclusion extends and generalizes related conclusions derived for the standard decaying dark matter model.
 
 \end{abstract}
\maketitle

\section{Introduction}

Within the standard cosmological model, our Universe comprises cold dark matter (CDM), dark energy (DE) $\Lambda$, a small amount of baryonic matter, and radiation in the form of photons as well as massive neutrinos. The six parameters of this $\Lambda$CDM model have been constrained to very high ($\sim$percent) precision by cosmological datasets such as the anisotropy power spectrum of the cosmic microwave background (CMB) measured by the \textit{Planck} satellite~\cite{2020A&A...641A...1P}. However, as we have obtained ever more precise constraints, tensions of varying significance between different datasets have emerged. One of the most well-known of these is the $H_0$ tension, the discrepancy of $\sim5\sigma$ between certain measurements of the expansion rate of the Universe today when inferred from different datasets. The $H_0$ tension is largely driven by the discrepancy in the local measurement of the SH0ES collaboration~\cite{2019NatAs...3..891V,2020PhRvD.101d3533K,2021CQGra..38o3001D,2021A&ARv..29....9S} and the model-dependent inference from the CMB (e.g., as constrained with \textit{Planck} data) and/or large-scale structure data; it should be noted that many local measurements of $H_0$ are consistent with both \textit{Planck} and SH0ES, albeit with larger error bars than the SH0ES measurement~\cite{2019ApJ...882...34F}. In the coming years, such independent measurements of $H_0$ are forecast to get more precise, hopefully deciding finally whether the $H_0$ tension indeed requires new physics.

At somewhat less significance ($\sim2-3\sigma$) is the $S_8$ tension, a tension in the amount of clustering of matter seen between many late-Universe datasets and the CMB data (see eg~\cite{2020A&A...633L..10T,2021A&A...645A.104A,2022PhRvD.105b3520A,2022PhRvD.105d3517P,2020JCAP...05..042I,2021JCAP...12..028K}). While this tension is not yet as strong as the $\sim 5\sigma$ tension seen in some probes of $H_0$, its consistency across several datasets is certainly intriguing.

These tensions could either be fluctuations (in the case of the less significant $S_8$ tension); caused by experimental or astrophysical systematic effects; or real hints to new physics. If we do take the tensions at face value, as indicative of new physics, a simultaneous resolution of the two would be very compelling. In this work we consider a scenario with the potential for achieving this goal: a model which modifies the expansion rate and structure growth of the Universe by positing that some portion of the CDM has converted to dark radiation (DR) after the release of the CMB. This leads to a lower density of CDM today than predicted by the $\Lambda$CDM model fit to the \textit{Planck} data, and as a result the DE  becomes dominant earlier and leads to more accelerated expansion---and thus, a higher value of $H_0$. Simultaneously, due to a) the decay of some of the CDM after recombination; and b) the free-streaming of the DR decay product (which suppresses clustering), this model can also lead a decreased matter power spectrum $P(k)$, and thus a lower $S_8$---a property that has led to this being suggested  as a potential simultaneous solution of the tensions~\cite{2018PhRvD..98b3543B,2021PhRvD.103l3528C}.

In Ref.~\cite{2018PhRvD..98b3543B}, a  phenomological, model-independent prescription for this conversion of CDM to DR was introduced; this includes (and generalizes) a decaying CDM (DCDM) scenario~(e.g.,~\cite{1985PhRvD..31.1212T,2014JCAP...12..028A,2016JCAP...08..036P})  in which some fraction of the CDM is unstable with a cosmological-scale lifetime $\tau$ and decays into DR. The amount of DCDM  has been constrained using CMB, BAO, and LSS data (e.g.,~\cite{2004PhRvL..93g1302I,2014JCAP...12..028A,2015JCAP...09..067E,2015PhRvD..92l3516A,2016PhRvD..94b3528C,2016JCAP...08..036P,2020JCAP...04..015E,2018PhRvD..97h3508C,2020JCAP...01..045X,2021JCAP...05..017N,Alvi:2022aam,Simon:2022ftd}),  and also investigated as a solution to the Hubble and/or $S_8$ tensions~(e.g.,~\cite{2015JCAP...09..067E,2020JCAP...07..026P,2020MNRAS.497.1757H,2022arXiv220304818A,Davari:2022uwd}).

The model of Ref.~\cite{2018PhRvD..98b3543B} allows for a time-dependent conversion of some fraction of the CDM to DR, occurring at an arbitrary cosmological time (set by a parameter of the theory); this should be contrasted with the exponential conversion of CDM to DR at the cosmological time corresponding to its lifetime $\tau$ in the DCDM scenario. It was noted explicitly in Ref.~\cite{2018PhRvD..98b3543B} that this model can result in a higher $H_0$ as well as a lower value of $S_8$ than in $\Lambda$CDM and thus has the potential to allow for the simultaneous resolution of the $H_0$ tension and the $S_8$ tension.

In this work, we constrain this DM$\rightarrow$DR model, and compare it to $\Lambda$CDM using a Bayesian approach to investigate if it can indeed solve the Hubble or $S_8$ tensions. We find that, for CMB data, it is \textit{not preferred} over $\Lambda$CDM, and that even when the SH0ES $H_0$ constraint is included in the analysis, the amount of CDM that converts to DR is constrained such that $H_0$ does not significantly increase relative to $\Lambda$CDM, while $S_8$ also remains nearly the same. We conclude that a model in which some fraction of the DM has converted to DR since recombination will not solve the cosmological concordance problem, unless other modifications are also considered, such as changes to the equation(s) of state of the species involved or additional (self)-interactions. In the course of our investigation, we explain the origin of these constraints in detail and correct various aspects of earlier implementations of this scenario.  Our modified Boltzmann code is publicly available\footnote{\url{https://github.com/fmccarthy/class_DMDR}}.

An outline of this paper is as follows. In Section~\ref{sec:tensions}, we discuss the $H_0$ and $S_8$ tensions. In Section~\ref{sec:theory}, we outline the theory of the  DM$\rightarrow$DR model, including the modifications to the homogeneous Universe and the perturbation structure. In Section~\ref{sec:data}, we describe the data products and likelihoods used in our analysis. In Section~\ref{sec:results}, we present our results. In Section~\ref{sec:conclusion}, we  discuss our results and conclude.

\section{Cosmological tensions}\label{sec:tensions}

\subsection{The $H_0$ tension}
Assuming $\Lambda$CDM, the \textit{Planck} CMB data predict $H_0 = 67.4\pm0.5 \, \mathrm{km/s/Mpc}$~\cite{2020A&A...641A...6P}. This is derived from the direct measurement of the angular size of the acoustic scale in the CMB power spectrum. Some local measurements, which measure $H_0$ directly by constructing a distance-redshift relation (the [cosmic] ``distance ladder''), are in tension with this result, e.g., the most recent measurement from the SH0ES collaboration, $H_0=73.04\pm1.04 \,\mathrm{km/s/Mpc}$~\cite{2021arXiv211204510R}.

If this tension is not due to experimental or astrophysical systematics, one of these inferences is incorrect, and the tension can be taken as an indicator of new physics. The direct measurement is (in principle) model-independent, and thus we should address the modeling that predicts $H_0$ from the directly-constrained acoustic scale in the CMB.  This calculation of $H_0$ relies on our model of the expansion history of the Universe since the CMB was released in the early Universe (at ``recombination''), when the Universe was very young; and our model of the sound horizon at recombination. Solutions to the Hubble tension must modify (at least) one of these models, while remaining consistent with the \textit{Planck} data. For a recent review of the proposed models to alleviate this tension, see~\cite{2022PhR...984....1S}.

\subsubsection*{Direct measurements of $H_0$: the cosmic distance ladder}

We can measure $H_0$ today by directly measuring the apparent recession velocity and distance to distant objects. While velocity can be measured directly by measuring the redshift of spectra of objects, distance requires the use of a ``standard candle'' of known intrinsic brightness along with the distance-luminosity relation. Type Ia Supernovae (SNe) can be used as a standardizable candle to measure $H_0$; however, the normalization of their brightness is not known absolutely, and so they can only constrain the relative evolution of cosmological distance, $H(z)/H_0$. To constrain their intrinsic brightness, we need to know the absolute distance to some of the SNe; we measure this by using other standard candles, such as cepheids, which are known to obey a tight period-luminosity relation~\cite{1912HarCi.173....1L}. In turn, the cepheid intrinsic brightness is measured by taking parallax measurements of the nearest cepheids, in particular those in nearby galaxies. Thus we have the cosmic distance ladder: the parallax measurements of nearby cepheids are used to calibrate the more distant cepheids, which in turn are used to calibrate the nearby SNe; using this calibration, these and the more distance SNe are used to measure $H_0$. 

The SH0ES collaboration uses this approach to measure $H_0$ directly as $H_0=73.04\pm1.04 \,\mathrm{km/s/Mpc}$~\cite{2021arXiv211204510R}. Other methods include using tip of the red giant branch (TRGB) stars instead of cepheids to calibrate the SNe; these measurements are in less tension with \textit{Planck}, finding $H_0=69.8\pm1.9$ km/s/Mpc~\cite{2019ApJ...882...34F}.

\subsubsection*{Inference of $H_0$ from the CMB}

We infer $H_0$ from the angular scale $\theta_s$ imprinted on the CMB by baryonic acoustic oscillations (BAOs). $\theta_s$  is a projection of the physical sound horizon at recombination $r_s(z^\star)$, according to
\be
\theta_s = \frac{r_s(z^\star)}{D_A(z^\star)},\label{theta_rs_DA}
\ee
where $D_A(z^\star)$ is the comoving angular diameter distance to the surface where the CMB was released at redshift $z^\star$ (the ``surface of last scattering'').

The sound horizon $r_s(z^\star)$, the distance a sound wave could travel in the time between the beginning of the Universe and recombination, is given by the integral over comoving distance multiplied by the sound speed $c_s(z)$:
\be
r_s(z^\star) = \int_{z^\star} ^\infty\frac{dz}{H(z)}c_s(z);
\ee
 the comoving distance is given by\footnote{We work in units with the speed of light $c=1$.}
\be
D_A(z^\star) = \int _0^{z^\star}\frac{dz}{H(z)}.
\ee
In these distance integrals, $H(z)$ accounts for the geometry of the expanding Universe.  Its form depends on the density of the Universe via the Friedmann equation:
\be
H(z) = {\sqrt{\frac{8\pi G}{3}\rho(z)}}\label{H_general}
\ee
 where $\rho(z)$ is the energy density of the Universe at redshift $z$. Within $\Lambda$CDM, {the form of} $\rho(z)$ is specified explicitly, and thus so is the $z$-dependence of $H(z)$:
 \begin{align}
 \rho(z)^{\Lambda CDM}=&\rho_m(z)+\rho_\gamma(z)+\rho_\nu(z)+\rho_\Lambda\label{rho_z}\\
 =&\rho_m^0(1+z)^3+\rho_\gamma^0(1+z)^4+\rho_\nu(z)+\rho_\Lambda,
 \end{align}
 where the subscripts $\{m,\gamma,\nu,\Lambda\}$ refer to the components of the Universe within $\Lambda$CDM: matter $m$  (including CDM and baryons); radiation $\gamma$ (including photons and massless neutrinos); massive neutrinos $\nu$; and the cosmological constant $\Lambda$, respectively; the values of $\rho_i$ must be measured to fully characterize $\rho(z)^{\Lambda CDM}$.

 $\rho_m^0$ and $\rho_\gamma^0$ are the densities of matter and radiation today; $\rho_m^0$ is constrained indirectly from the CMB, which most directly constrains $\rho_m(z\approx z^\star)$, by assuming the standard evolution of matter $\rho_m(z)\propto(1+z)^3$.  
 $\rho_\gamma^0$ is constrained from the monopole temperature of the CMB~\cite{Fixsen_2009}. $\rho_\nu(z)$, the evolution of the neutrino density, is also constrained from the CMB; the form of its evolution $\rho_\nu(z)$ depends on the neutrino mass but is specified within $\Lambda$CDM. The only remaining component of the density is $\rho_\Lambda$. However, the CMB directly constrains $\theta_s$; and the physics of the sound speed $c_s(z)$  are well understood within $\Lambda$CDM.  Thus, the CMB data along with Equation~\eqref{theta_rs_DA}  specify $\rho_\Lambda$; as such, $H(z)$, including its value today $H_0\equiv H(z=0)$, is fully specified by the CMB within $\Lambda$CDM, although it is useful to break the ``geometric degeneracy''~\cite{1999MNRAS.304...75E} in the fit to CMB data using an external probe of the matter density, such as BAO or CMB lensing data. \textit{Planck} finds $H_0 = 67.4\pm0.5 \, \mathrm{km/s/Mpc}$~\cite{2020A&A...641A...6P}; similar inferences which use the BAO scale of galaxy surveys (as opposed to the CMB) are in agreement with this, with the DES survey combined with BOSS BAO data and BBN data finding $H_0 = 67.4\pm1.2 \, \mathrm{km/s/Mpc}$~\cite{2018MNRAS.480.3879A}.

$H_0$ inferences using the cosmic ``inverse'' distance ladder (see, e.g.~\cite{2019MNRAS.486.2184M,2020MNRAS.495.2630C}), wherein an absolute SNIa luminosity calibration is determined directly from the BAO scale (by comparing directly luminosity and the BAO angular distance measurement at the same redshift), are also not in tension with \textit{Planck}, with~\cite{2020MNRAS.495.2630C} finding  $H_0 = 69.71 \pm 1.28$ for an analysis in which the SNe were calibrated from the BAO angular sound horizon (which itself was calibrated from the CMB angular sound horizon). Such methods disfavor models that modify cosmic evolution after recombination to attempt to increase $H_0$, as they depend only on the fact that the BAO scale is the same at $z \approx 1100$ and at low redshifts.  However, the error bars are large enough that some potential wiggle room remains.

To infer a different value of $H_0$, there are three options: modify the pre-recombination sound speed; modify $\rho(z)$ before recombination; or modify $\rho(z)$ after recombination (or some combination of these). In this work, we focus on the modification of $\rho(z)$ after recombination, in particular by modifying the evolution of the CDM density. We allow some component of the CDM  to convert into dark radiation (DR), and thus modify the form of $\rho_m(z)$ while also adding a new component $\rho_{DR}(z)$ to Equation~\eqref{rho_z}. This can lead to a different value of $\rho_\Lambda$, as well as a different value of $\rho_m^0$, a different $z$-evolution $H(z)$, and a different value of $H_0$ today.

\subsection{The $S_8$ tension}\label{sec:s8tension}

The $S_8$ parameter is defined as 
\be
S_8 \equiv \sigma_8\lb\frac{\Omega_m}{0.3}\rb^{0.5}.
\ee
 $\Omega_m \equiv \rho_m^0 / \rho_{cr}^0$ is the density of matter today as a fraction of the critical density $ \rho_{cr}^0$ and $\sigma_8$ measures the rms amplitude of linear matter density fluctuations over a sphere of radius $R=8 \, \mathrm{Mpc}/h$ at $z=0$:

\be
\lb\sigma_8 \rb^2= \frac{1}{2\pi^2}\int \frac{dk}{k}W^2(kR) k^3 P(k),
\ee
where $P(k)$ is the linear matter power spectrum today and $W(kR)$ is a spherical top-hat filter of radius $R=8\mathrm{Mpc}/h$.

There is a slight tension emerging between $S_8$ as measured from late-Universe datasets and indirecty inferred the CMB; i.e., by constraining the $\Lambda$CDM parameters from the CMB and calculating the resulting $S_8$.  In particular, weak lensing surveys such as KIDS measure $S_8=0.759\pm0.024$~\cite{2021A&A...645A.104A};  clustering surveys analyses such as BOSS also consistently find low $S_8$~\cite{2022PhRvD.105d3517P,2020JCAP...05..042I}.
DES measures $S_8 = 0.776\pm0.017$ from galaxy-galaxy lensing~\cite{2022PhRvD.105b3520A} (the combined analysis of the clustering of foreground galaxies and lensing of background galaxies). These numbers should be compared to the indirect \textit{Planck} constraint, $S_8=0.834\pm0.016$ from the primary CMB~\cite{2020A&A...641A...6P}. While CMB lensing from \textit{Planck} alone is not in tension with respect to the primary CMB constraints, its low-$z$ contribution, measured through cross-correlation with the unWISE galaxy sample (galaxies at redshifts at around $z\sim1-2$) gives $S_8=0.784\pm0.015$~\cite{2021JCAP...12..028K}, an interesting addition to the low-$S_8$ measurements due to its complementary systematics.

\section{Theory}\label{sec:theory}
\subsection{Background cosmology}
We consider $\Lambda$CDM modified by the addition of an extra dark matter (DM) component, such that the total background DM density evolves as~\cite{2018PhRvD..98b3543B}
\be
\rho_{DM}(a) = \frac{\rho_{DM}^0}{(a/a_0)^3}\lb1+\zeta\lb\frac{1-(a/a_0)^\kappa}{1+\lb \frac{a/a_0}{a_t}\rb^\kappa}\rb\rb\label{rhochia},
\ee 
where $\rho_{DM}^0$ is the total DM density today, $a$ is the scale factor, and $a_0$ is a reference scale factor.  This modification to $\Lambda$CDM is fully characterized by three parameters: $\zeta,\kappa$, and $a_t$. $\zeta$ describes the amount of DM that converts into DR---in particular, the comoving DM density decreases by a factor of $(1+\zeta)$ between $a\rightarrow 0$ and $a_0$;
$\kappa$ characterizes the rate of the DM$\rightarrow$DR conversion;  and $a_t$ sets the timescale for the conversion to occur. For transitions long before $a_0$ (i.e., for which $\lb a_0/a_t\rb^\kappa\ll a_0$), all of the DM remaining at $a_0$ evolves as normal and $\rho_{DM}(a)$ can be split neatly into a ``converting'' and a standard component (which decays as $a^{-3}$) by inspection. For transitions that are not complete at $a_0$ (or for transitions in the future relative to $a_0$), this is not the case, as some of the contribution to $\rho_{DM}^0$ will decay; however, it is possible to reparameterize Equation~\eqref{rhochia} by redefining the reference scale factor $a_0$ such that $\lb a_0/a_t\rb^\kappa\ll\ a_0$ in such a way that there is a well-defined split into the converting and standard component. In any case, the final comoving DM density ($a^3\rho_{DM}(a\rightarrow\infty)$) is given by $ \frac{\rho_{DM}^0}{(1/a_0)^3}\lb1-a_t^\kappa\zeta\rb$. The requirement that the DM density always be positive thus gives a constraint on the parameters: we require 
\be
\zeta\le\frac{1}{a_t^\kappa}.\label{physicality_condition}
\ee
This also allows for the case that \textit{all} DM is of the converting type and will eventually convert, in which case  the inequality in Equation~\eqref{physicality_condition} is saturated and $a^3\rho_{DM}(a\rightarrow\infty)=0$.

Hereafter, we will always take $a_0$ to be the scale factor today and set $a_0=1$. Note that, in this case, $\zeta$ describes the fraction of the original DM that has converted by today, but \textit{not} the total fraction of DM that will eventually convert, unless the transition is in the past: $\lb 1/a_t\rb^\kappa\ll 1$. In this parametrization it is evident that scenarios in which the transition is yet to begin are degenerate with $\Lambda$CDM as they demand $\zeta\rightarrow0$. 

We consider the case where the DM converts into a dark radiation (DR) particle, whose background energy density evolves as $a^{-4}$. The conservation of energy demands that
\be
\frac{1}{a^3}\frac{d}{dt}\lb a^3 \rho_{DM}\rb = -\frac{1}{a^4}\frac{d}{dt}\left( a^4 \rho_{DR}\right),
\ee
which allows us to explicitly write the DR energy density $\rho_{DR}$ as~\cite{2018PhRvD..98b3543B}
\begin{align}
&\rho_{DR}(a) = \zeta\frac{\rho_{DM}^0}{a^3}\frac{\lb 1+a_t^\kappa\rb}{\lb a^\kappa+a_t^\kappa\rb}\times\nonumber\\
&\,\,\left(\lb a^\kappa+a_t^\kappa\rb {}_2F_1\left[1,\frac{1}{\kappa};1+\frac{1}{\kappa};-\lb\frac{a}{a_t}\rb^\kappa\right]-a_t^\kappa\right)\,, \label{rhoDR}
\end{align}
 where $_2F_1(b,c;d;z)$ is the hypergeometric function.

The ansatz in Equation~\ref{rhochia} encompasses and generalizes the standard decaying DM model, in which a sub-component of the DM exponentially decays with lifetime $\tau$.  Such a model is accurately captured by setting $\kappa = 2$ in Equation~\ref{rhochia} and setting $a_t$ such that $H(a_t) \approx \Gamma$, where $\Gamma$ is the DM decay rate.  As pointed out in Ref.~\cite{2018PhRvD..98b3543B}, a model in which a sub-component of the DM undergoes Sommerfeld-enhanced annihilation can be accurately represented by setting $\kappa=1$.  This approach thus naturally encompasses a wide range of possible scenarios in which DM converts to DR, in a relatively model-independent manner.

\subsubsection*{Impact on the expansion of the Universe}

\begin{figure*}
\includegraphics[width=\columnwidth]{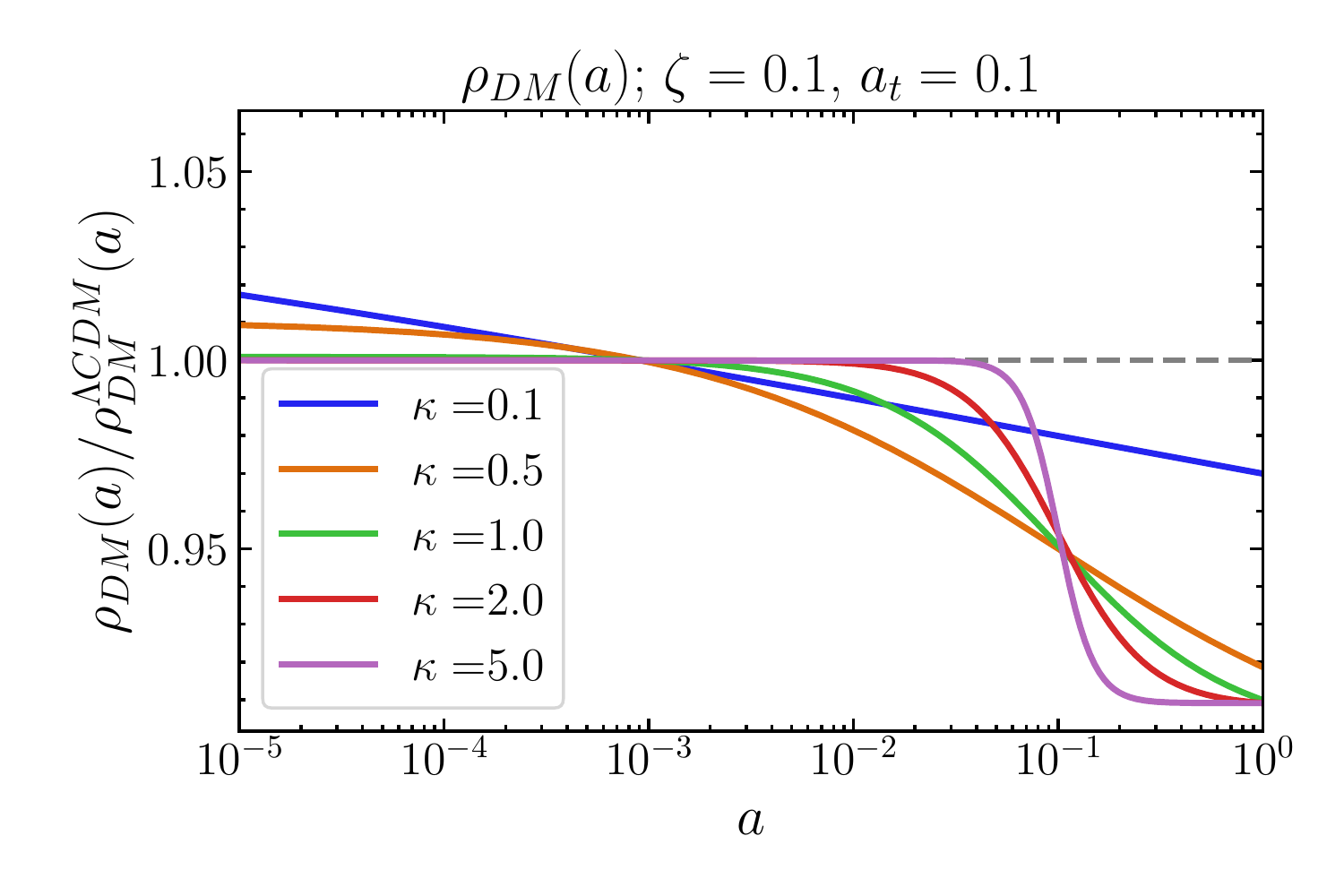}
\includegraphics[width=\columnwidth]{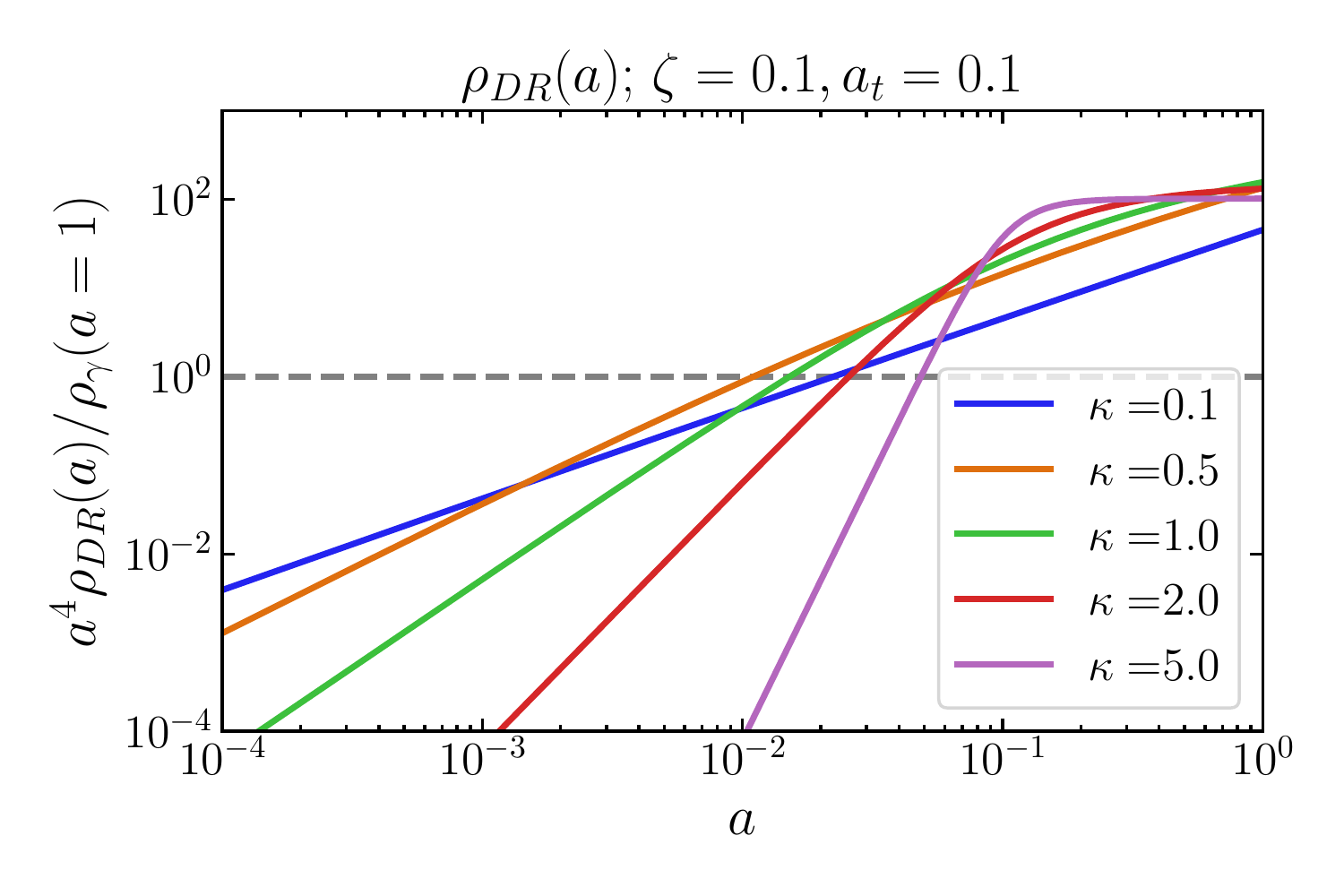}
\caption{\textit{Left:} The evolution of the CDM density in the DM$\rightarrow$DR model for various values of the conversion rate $\kappa$ (with $\zeta = 0.1$ and $a_t = 0.1$ fixed), compared to that of $\Lambda$CDM. The parameters are chosen such that $\theta_s$, the angular scale of the CMB peaks, is the same for each case; the density of CDM $\Omega_{CDM} h^2$ is modified such that the CDM densities are equal in the two models at $z \approx 1100$ when the CMB was released; this is evident in the $\rho_{CDM}$ plot where we see that the curves all intersect at $a \approx 10^{-3}$. \textit{Right:} The evolution of the DR density, with the evolution of the standard photon density also shown (by the gray dashed line); this is unchanged between the $\Lambda$CDM and DM$\rightarrow$DR models, with the photon density always evolving as $a^{-4}$ and with its value set by the temperature of the CMB.}\label{fig:density_evolution}
\includegraphics[width=\columnwidth]{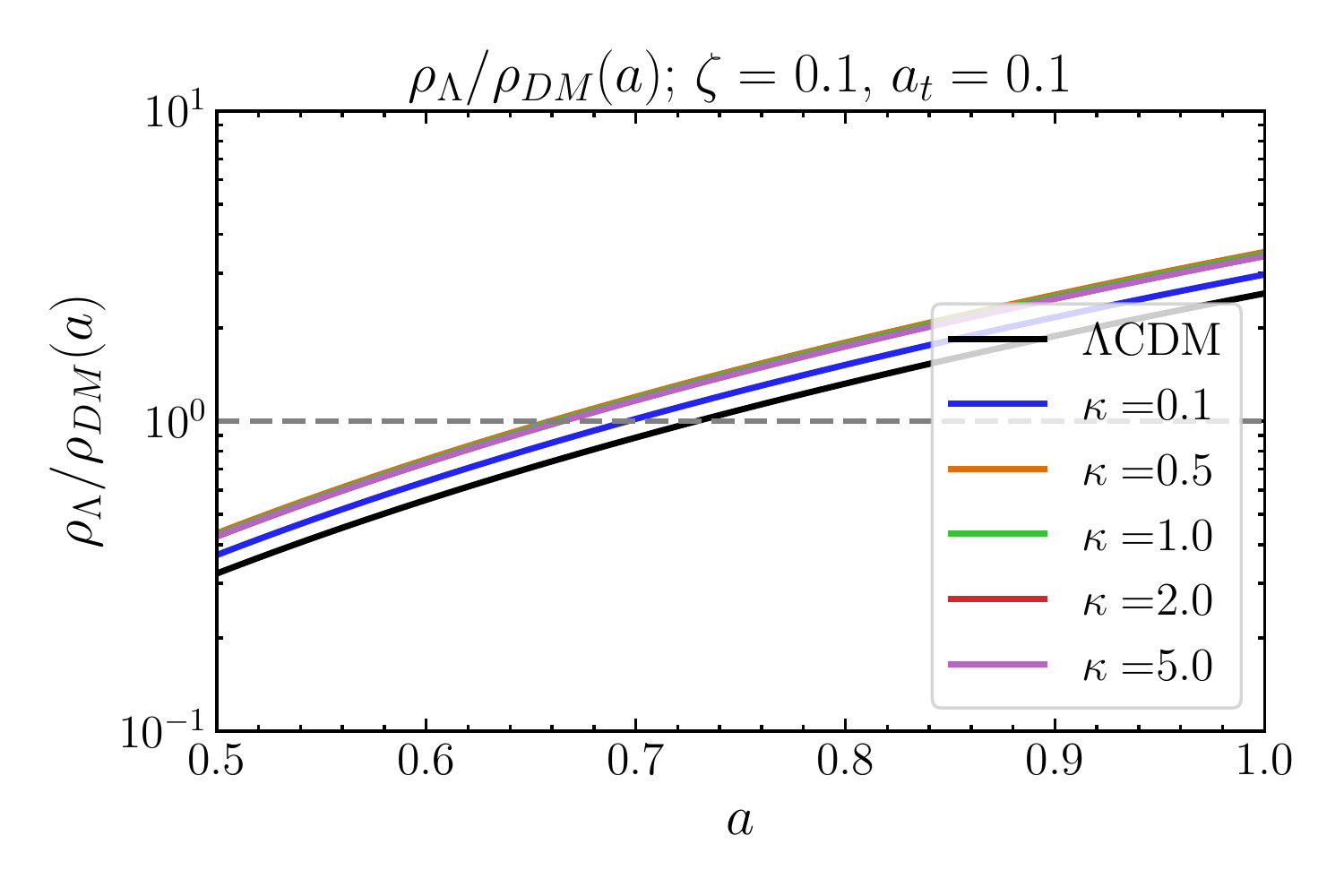}
\includegraphics[width=\columnwidth]{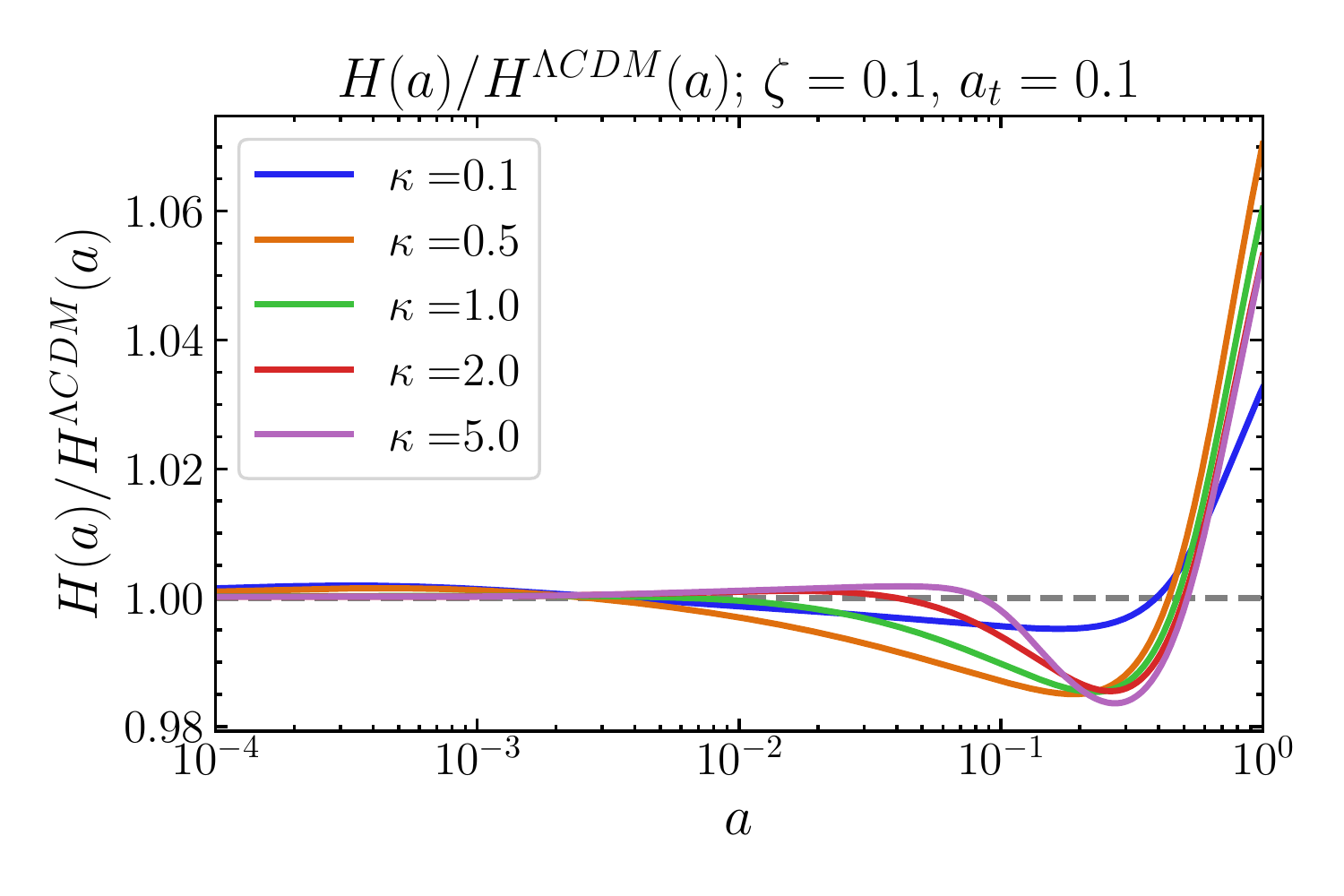}
\caption{\textit{Left:} The (constant) DE density $\rho_\Lambda$ divided by the total CDM density $\rho_{CDM}$. We see that the departures from $\Lambda$CDM increase the redshift of matter-$\Lambda$ equality such that $\Lambda$ becomes dominant in the DM$\rightarrow$DR model earlier than in $\Lambda$CDM. This leads to a higher value of $H_0$ today, as shown in the right panel. \textit{Right:} The evolution of the Hubble parameter $H(z)$ in the DM$\rightarrow$DR model compared to $\Lambda$CDM.}\label{fig:H_evolution}
\end{figure*}

This modification to the evolution of the background density of the Universe directly changes the evolution of the Hubble rate $H(a)$. In order to compare with $\Lambda$CDM, we must think about what parameters should remain fixed. In our $\Lambda$CDM plots in Figures~\ref{fig:density_evolution} and~\ref{fig:H_evolution}, we fix the background cosmological parameters to the best-fit values from the \textit{Planck} fit to the CMB alone (TT-EE-TE): thus for the $\Lambda$CDM case we take  $\{100\theta_s=1.040909,\Omega_bh^2=0.022383,\Omega_{CDM}h^2=0.12011\}$, where $\theta_s$ is the angular size of the acoustic scale at last scattering, $\Omega_bh^2$ is the physical density of baryons today, and $\Omega_{CDM}h^2$ is the  physical density  of CDM today (these quantities are directly constrained by the CMB).

For the modified DM case, we must consider that the CMB does not directly constrain the density of CDM today, but instead the density of CDM when it was released at $z=z^\star \approx 1100$, the redshift of the surface of last scattering. Thus we modify $\Omega_{CDM}h^2$ in the DM$\rightarrow$DR plots to demand that the matter density at the redshift of last scattering matches the constraint from the CMB; this results in the relation
\begin{eqnarray}
& \left(\Omega_{CDM} h^2\right)_{\mathrm{\Lambda CDM}} = \nonumber \\ 
  & \left(\Omega_{CDM} h^2\right)_{\mathrm{DM\rightarrow DR}}\lb1+\zeta\frac{1-a_\star^\kappa}{1+\lb\frac{a_\star}{a_t}\rb^\kappa}\rb
\end{eqnarray}
where $a_\star=\frac{1}{1+z^\star}$ was the scale factor at the time of last scattering. Note that $\left(\Omega_{CDM} h^2\right)_{\mathrm{DM\rightarrow DR}}$ includes the density both of the converting part and the non-converting part of the DM.

 The evolution of the resulting DM and DR densities are  shown in Figure~\ref{fig:density_evolution}, for some choices of the DM$\rightarrow$DR parameters. Note that because we hold $\theta_s$ fixed in each case, it is not immediately straightforward to calculate the evolution of these densities directly, as one must deduce $H_0$ (more generally $H(z)$, in particular by finding the dark energy density $\rho_\Lambda$ required to make the Universe flat) appropriately. In the plots in Figures~\ref{fig:density_evolution} and~\ref{fig:H_evolution}, we have deduced $H_0$ using the ``shooting'' method implemented in \texttt{CLASS}~\cite{2011JCAP...07..034B}.
 
In Figure~\ref{fig:H_evolution}, we show on the left how these density evolutions lead to an earlier redshift of $\Lambda$-matter equality, and thus result in a higher value of $H_0$ today, as is shown on the right.

\subsection{Perturbative cosmology}

The perturbations to the homogeneous background, which give rise to the CMB and the clustering of matter today, are evolved with the linearized perturbed Einstein--Boltzmann equations. This is done numerically, usually with an Einstein--Boltzmann solver such as \texttt{CLASS}\footnote{\href{https://lesgourg.github.io/class_public/class.html}{https://lesgourg.github.io/class\_public/class.html}}~\cite{2011JCAP...07..034B} or \texttt{CAMB}\footnote{\href{https://camb.info/}{https://camb.info/}}~\cite{Lewis:1999bs}. In our implementation of the DM$\rightarrow$DR dark matter model, we modify \texttt{CLASS}.

Both the DM and the DR perturbations must be evolved correctly, even though the DR perturbations cannot be detected directly: they interact gravitationally  with the perturbations to the Universe's spacetime metric, and thus indirectly with the measureable perturbations of interest, in particular the DM perturbations and the perturbations to the photon fluid (which we observe as the CMB). Following Ref.~\cite{1995ApJ...455....7M}, in this Subsection we present the Boltzmann equations for the DM and DR.

Formally, the fluids obey the Boltzmann equation
\be
\frac{df_i}{dt}=\mathcal{Q}_i
\ee
where $f_i$ is the distribution function of either DM or DR, and $\mathcal{Q}_i$ is the appropriate collision term. For the system as a whole, there are no collisions (i.e., $\sum_i\mathcal{Q}_i=0$) and so $\mathcal{Q}_{DR}=-\mathcal{Q}_{DM}\equiv\mathcal {Q}$. The 0th-order, momentum-integrated Boltzmann equation is
\be
\frac{1}{a^3}\frac{d}{dt}\lb a^3 \rho_{DM}^{(0)}\rb=-\frac{1}{a^4}\frac{d}{dt}\lb a^4 \rho_{DR}^{(0)}\rb=-\mathcal{Q}^{(0)},
\ee
where superscript $^{(0)}$ refers to 0th-order (background) quantities; and so we see (from comparison with the derivative of Equation~\eqref{rhochia}) that the 0th-order collision term is
\be
\mathcal{Q}^{(0)}=\frac{H \kappa \rho_{DM}^0\zeta}{a^3}\lb\frac{a^\kappa+\lb\frac{a}{a_t}\rb^\kappa}{\lb1+\lb\frac{a}{a_t}\rb^\kappa\rb^2}\rb .
\ee
The perturbations are evolved with the 1st-order Boltzmann equation; to evolve these we need to specify the perturbation to $\mathcal{Q}$, ie $\mathcal{Q}^{(1)}$. We follow the ``minimal option'' of Ref.~\cite{2018PhRvD..98b3543B}: 
\be
\mathcal{Q}^{(1)} = \mathcal{Q}^{(0)}\delta_{DM}\label{minimal_option}
\ee
where $\delta_{DM}\equiv\frac{\delta \rho_{DM}}{\rho_{DM}{}^{(0)}}$ is the dimensionless perturbation to the DM density, with $\delta \rho_{DM}\equiv\rho_{DM}^{(1)}$ the dimensionful perturbation.  The exact specification of $\mathcal{Q}^{(1)}$ is model-dependent, but any change from the ansatz in Equation~\eqref{minimal_option} must be proportional to $\mathcal{Q}^{(0)}$, as emphasized in Ref.~\cite{2018PhRvD..98b3543B}.  Since this is already tightly constrained (see below) solely by the evolution of background densities, any correction to this assumption will have a negligible impact on our results.  It is also worth noting that in the standard decaying DM scenario Equation~\eqref{minimal_option} is exact, and even in a Sommerfeld-enhanced annihilation scenario the corrections to it are negligible~\cite{2018PhRvD..98b3543B}.

In the synchronous gauge, the perturbed FRW (Friedmann--Robertson--Walker) metric is
\be
ds^2 = g_{\mu\nu} dx^\mu dx^\nu = a^2 \lb-d\tau^2+\lb\delta_{ij}+h_{ij}\rb dx^i dx^j\rb,
\ee
where $\tau$ is conformal time; $\delta_{ij}$ is the Kronecker delta in three (spatial) dimensions; and $h_{ij}$ are the synchronous metric perturbations (with three-dimensional trace $h\equiv \sum_ih_{ii}$). In this gauge the Boltzmann equations for the DM$\rightarrow$DR model are\footnote{Due to the ``minimal'' assumption for $\mathcal Q$, the Boltzmann equations for the perturbations to the component of the DM that converts into DR coincide with those for the perturbations to the component that undergoes standard evolution, and so we retain the very general subscript $_{DM}$ here.}:
\begin{align}
\delta^\prime_{DM} = -\frac{h^\prime}{2};\\
\theta^\prime _{DM} = - \mathcal H \theta_{DM},
\end{align}
   with prime ($^\prime$) denoting differentiation with respect to conformal time and $\mathcal H\equiv\frac{a^\prime}{a}$.

The DR field is defined in terms of its perturbed phase-space distribution
\be
f_{DR}(\vec x, \vec p,  \tau) = f_{DR}^{(0)}(p,\tau)\lb1+\Psi_{DR}\lb\vec x, \vec p, \tau\rb\rb
\ee
where $f_{DR}^{(0)}(p,\tau)$ is the background phase-space distribution and $\Psi_{DR}\lb\vec x, \vec p, \tau\rb$ is its perturbation. $\Psi_{DR}$ is expanded over the Legendre polynomials $\mathcal P_\ell(\mu)$ in moments according to
\be
\Psi_{DR,\ell}(\tau,  k, p) = \frac{1}{\lb-i\rb^\ell}\int ^1 _{-1}\frac{d\mu}{2}P_\ell(\mu) \Psi\lb\vec k, \vec p, \tau\rb
\ee
where $\mu\equiv \hat p \cdot \hat k$ is the angle between the wavenumber $\vec k$ (after going to Fourier space) and the DR momentum $\vec p$. Following Ref.~\cite{2016JCAP...08..036P}, we consider the rescaled momentum-integrated quantities $F_{DR,\ell}$ where
\be
F_{DR,\ell}\equiv r_{DR}(a)\frac{\int dp \, p^3 f_{DR}^{(0)}\Psi_{DR,\ell}(\tau,  k, p) }{\int dp \, p^3 f_{DR}^{(0)}}
\ee
with the background quantity $r_{DR}(a)$ given by
\be
r_{DR}(a)\equiv\frac{ \rho^{(0)}_{DR}(a)a^4}{\rho_{cr}^0}
\ee
where $ \rho^{(0)}_{DR}(a)$ is the background DR density and $\rho_{cr}^0$ is the critical density today; henceforward, we denote $r_{DR}(a)$ simply as $r_{DR}$, for concision.
The conformal time derivative of $r_{DR}$ is
\be
r_{DR}^\prime =r_{DR} \frac{ a \mathcal Q}{\rho^{(0)}_{DR}}.
\ee

The multipole moments of $F_{DR}$ can be related to the standard DR overdensity $\delta_{DR}$, velocity divergence $\theta_{DR}$, and shear stress $\sigma_{DR}$~\cite{1995ApJ...455....7M}:
\begin{align}
\delta_{DR}=\frac{F_{DR,0}}{r_{DR}};\\
\theta_{DR} = \frac{3k}{4r_{DR}}F_{DR,1};\\
\sigma_{DR} = \frac{F_{DR,2}}{2 r_{DR}}.
\end{align}

The Boltzmann equations for the moments $F_{DR,\ell}$ are:
\begin{align}
F^\prime_{DR,0} = - k F_{DR,1}-\frac{4}{3}r_{DR}\frac{h^\prime}{2}+r^\prime _{DR}
\delta_{DM};\label{DR_boltzmann0}\\
F^\prime_{DR,1} = \frac{k}{3}F_{DR,0}-\frac{2k}{3}F_{DR,2}+\frac{r^\prime_{DR}}{k}\theta_{DM};\\
F^\prime_{DR,\ell>1}=\frac{k}{2\ell+1}\lb\ell F_{DR,\ell-1}-\lb\ell+1\rb F_{DR,\ell+1}\rb.\label{DR_boltzmannell}
\end{align}

This system of equations is exactly equivalent to those for the standard decaying dark matter scenario presented in~\cite{2016JCAP...08..036P}, where our collision term $\mathcal{Q}$ can be related directly to a time-dependent inverse-lifetime of the decaying dark matter $\Gamma$ according to
\be
\Gamma(a) = \frac{\mathcal{Q}}{\rho_{\mathrm{DM\rightarrow DR}}(a)} \,, \label{Gamma_Q}
\ee
where $\rho_{\mathrm{DM\rightarrow DR}}(a)$ is the background density of the converting DM component. Explicitly,
\be
\Gamma(a) = \kappa H(a) \lb\frac{\lb\frac{a}{a_t}\rb^\kappa+a^\kappa}{\lb1-a^\kappa\rb\lb1+\lb\frac{a}{a_t}\rb^\kappa\rb}\rb.\label{Gamma_a}
\ee

Note that Equation~\ref{Gamma_a} diverges as $a \rightarrow 1$.  Numerically, we regulate this divergence by placing a ceiling on the ratio $\Gamma(a)/H(a)$, such that its value never exceeds 100.  We verify that in all regimes of interest studied in this paper, this choice has no impact on our results.

The equations derived here agree with those presented in Ref.~\cite{2021PhRvD.103l3528C}, which also explored the DM$\rightarrow$DR modification to $\Lambda$CDM, with a focus on the Dark Energy Survey (DES) data.  Note, however, that our DR equations are slightly different from those in Ref.~\cite{2018PhRvD..98b3543B}, which derived the DR equations by treating the DR as an imperfect fluid with anisotropic stress equal to that of the background photon fluid. Thus the equations of the higher moments of the DR Boltzmann expansion were adopted directly from the photon equations, an approximation which is not correct as the background DR density does not evolve as $a^{-4}$ (see Equation~\ref{rhoDR}).

\subsection{Numerical implementation}

Our scenario can be thought of as a generalization of the standard decaying DM model which allows for some time-dependent decay rate.  We implement this extended ``converting DM'' scenario in the Einstein-Boltzmann solver \texttt{CLASS},\footnote{\url{http://class-code.net}} by directly modifying  the standard decaying DM implementation to allow for a time-dependent $\Gamma(a)$ according to Equation~\eqref{Gamma_a}.  We release this modified version of \texttt{CLASS} publicly as \texttt{CLASS\_DMDR}\footnote{\url{https://github.com/fmccarthy/class_DMDR}} .  We have validated our Boltzmann code extensively by comparing to the publicly-available modification of \texttt{CAMB} that was released with Ref.~\cite{2021PhRvD.103l3528C}.  The two codes agree to 5\% precision in $\Delta C_\ell\equiv C_\ell-C_\ell^{\Lambda CDM}$ in the tests that we have run (note that this corresponds to a closer agreement of $< 0.5\%$ agreement in the raw $C_\ell$, as $\Delta C_\ell$ is on the order of $1-10\%$ of $C_\ell$).

Calculation of the background DR density requires evaluating a hypergeometric function in Equation~\eqref{rhoDR}, in particular
\begin{equation}
_2F_1\left[1,\frac{1}{\kappa};1+\frac{1}{\kappa};-\lb\frac{a}{a_t}\rb^\kappa\right].
\end{equation}
We calculate this with the GSL function \texttt{gsl\_sf\_hyperg}~\cite{gough2009gnu}. We discuss some subtleties of the implementation in Appendix~\ref{app:hypergeometric}.

\subsection{Effects on the CMB and matter power spectra}

\begin{figure}
\includegraphics[width=\columnwidth]{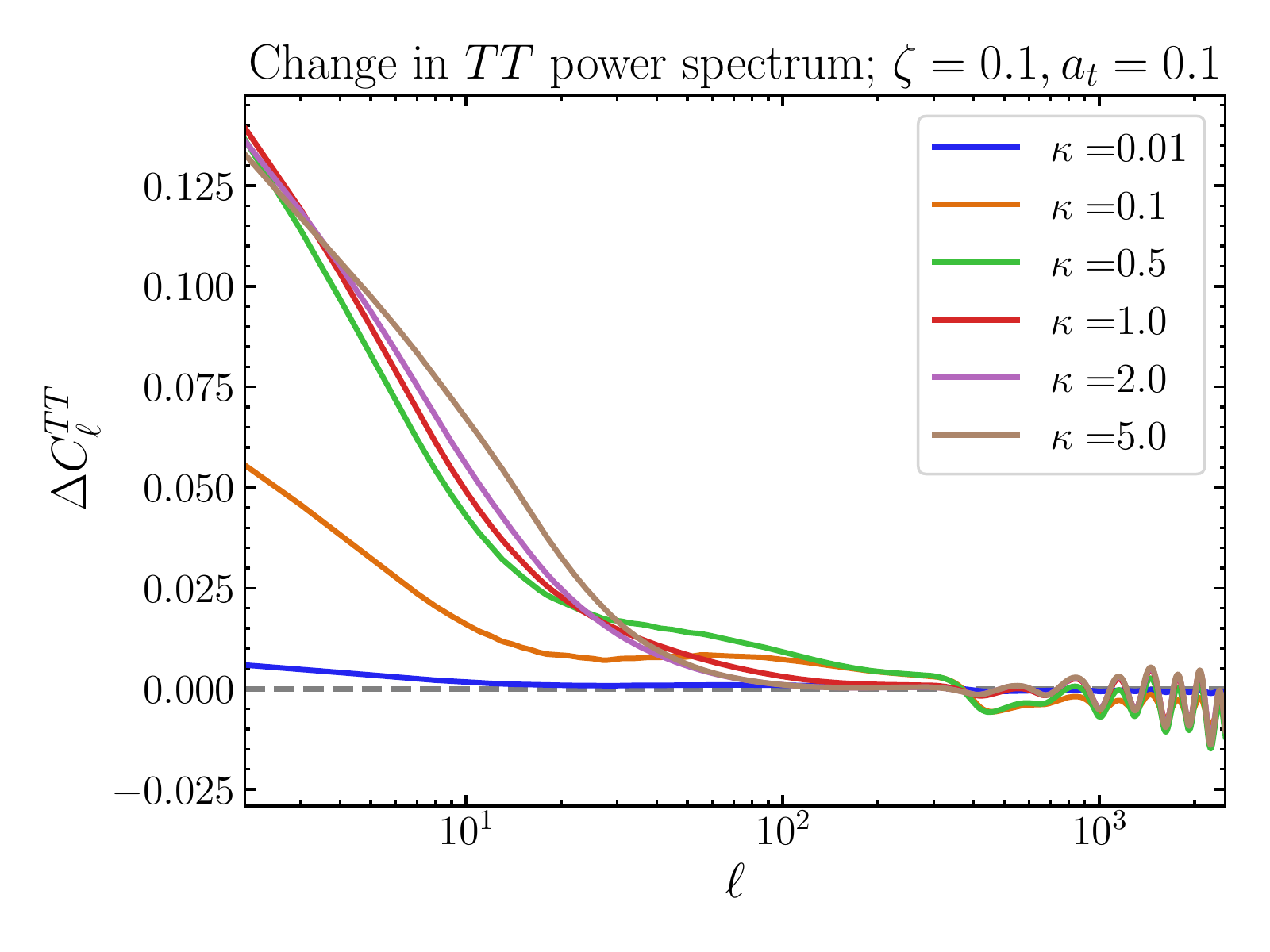}
\includegraphics[width=\columnwidth]{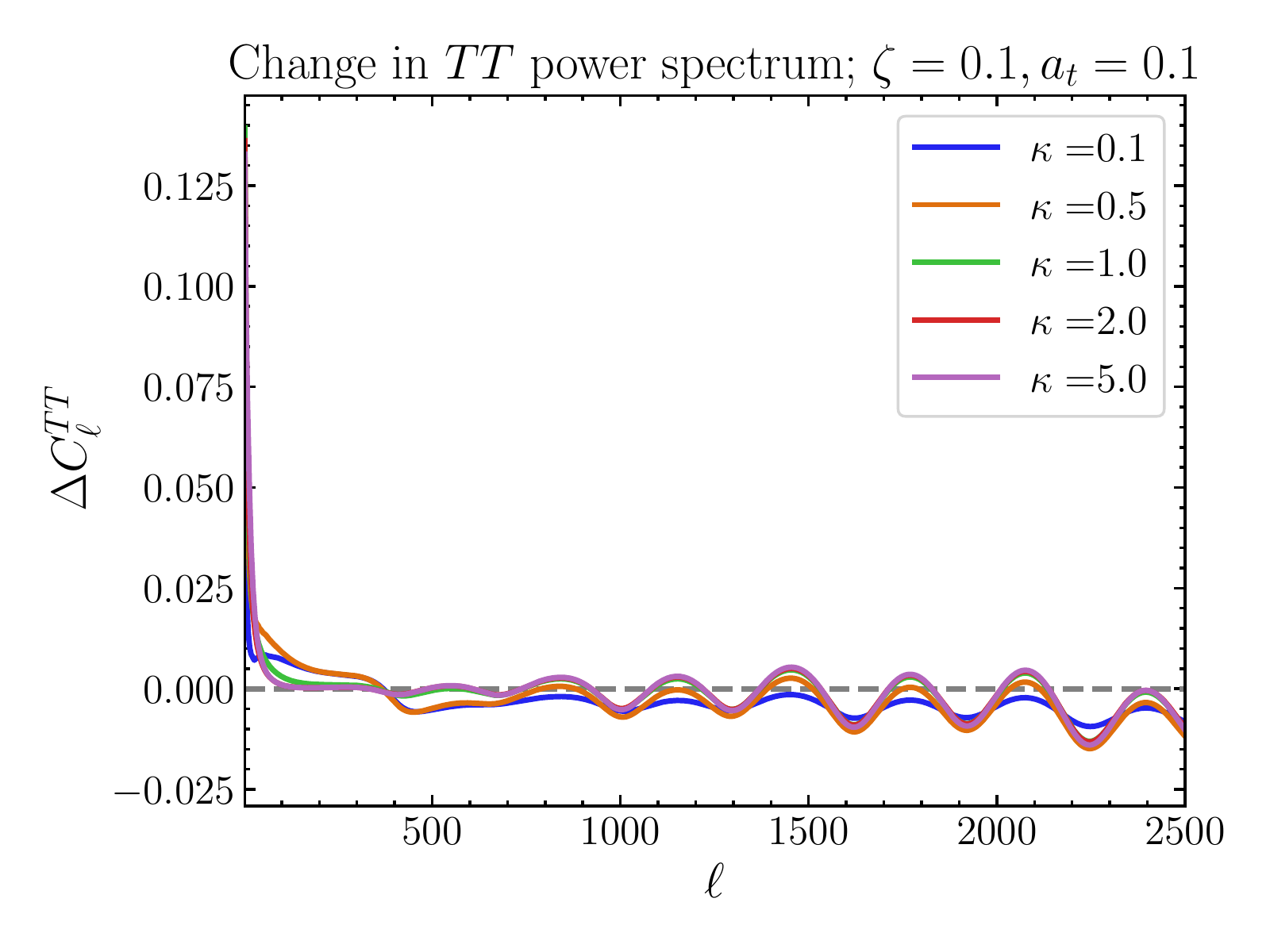}

\includegraphics[width=\columnwidth]{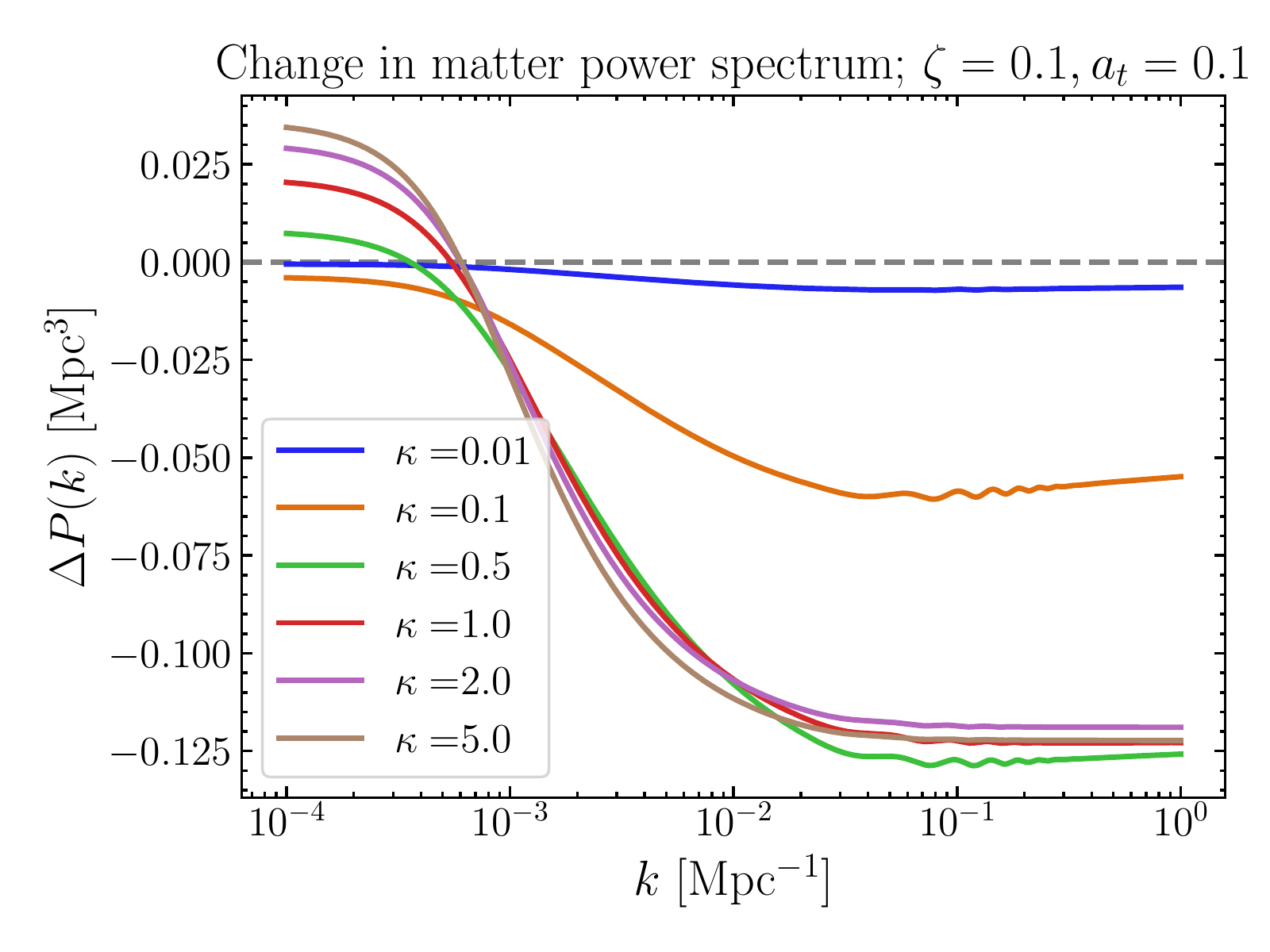}
\caption{Induced changes in the CMB temperature power spectrum (\textit{top} and \textit{middle}) and the linear matter power spectrum $P(k)$ (\textit{bottom}). In each case we define the fractional change $\Delta X \equiv\frac{X-X^{\Lambda CDM}}{X^{\Lambda CDM}}$, where $X$ is either $C_\ell^{TT}$ or $P(k)$. The biggest changes induced in $C_\ell^{TT}$ are due to the amplification of the late-time ISW signal, an effect which is mostly noticeable at low $\ell$, and due to the reduced impact of CMB lensing, an effect that is mostly noticeable at high $\ell$ (the characteristic peak-smearing due to lensing is lessened, and there is less power shifted from large to small scales). In the matter power spectrum, we see a suppression of power on scales $k \gtrsim 10^{-3} \,\, \mathrm{Mpc}^{-1}$. Note that the top and middle plots are identical except for the respectively logarithmic and linear $x$-axes scales. }\label{fig:deltacell}
\end{figure}

We show in Figure~\ref{fig:deltacell} the induced change to the CMB temperature power spectrum; we define the fractional change $\Delta C_\ell^{TT}\equiv\frac{C_\ell-C_\ell^{\Lambda CDM}}{C_\ell^{\Lambda CDM}}$. We have fixed the cosmological parameters similarly to how we did in Figures~\ref{fig:density_evolution} and~\ref{fig:H_evolution}, with the addition of the parameters relevant for $C_\ell^{TT}$ $\{A_s=2.1006\times 10^{-9},n_s=0.96605,\tau=0.0543\}$, where $A_s$ is the amplitude of scalar fluctuations (defined as standard with a pivot scale $k=0.05 \,\, \mathrm{Mpc^{-1}}$); $n_s$ is the scalar spectral index; and $\tau$ is the optical depth to reionization. These values are also determined from the \textit{Planck} analysis of $TT+TE+EE$ spectra.

In Figure~\ref{fig:deltacell}, one of the most evident changes from $\Lambda$CDM is an increase in the power spectrum at low $\ell$. This can be understood as an amplification of the late-time integrated Sachs--Wolfe (ISW) signal, which is sourced by the photons travelling through a time-dependent gravitational potential. In a purely CDM-dominated Universe, with $\rho\propto a^{-3}$, gravitational potentials are constant and there is no ISW effect. The late-time ISW effect is larger in the DM$\rightarrow$DR model due to the additional dark energy required to preserve flatness (see Figure~\ref{fig:H_evolution}) and, less significantly, due to the  presence of the DR and the modified evolution of the DM.  Since the late-time ISW is largest at low-$\ell$ and very subdominant at high-$\ell$, this results in the largest increase in $C_\ell^{TT}$ being at low $\ell$.

The other noticeable changes in the CMB temperature power spectrum in Figure~\ref{fig:deltacell} are due to the reduced effect of CMB lensing in the DM$\rightarrow$DR model, which follows from the decrease in the DM density (and increase in the DE density). It is well known that CMB lensing smears the peaks and troughs of the CMB power spectrum, and transfers power from large to small scales (e.g.,~\cite{2006PhR...429....1L}). In the DM$\rightarrow$DR scenario, there is less CMB lensing, and thus peak-smearing is reduced, resulting in a coherent raising of the peaks and lowering of the troughs in this model as compared to $\Lambda$CDM.  In addition, there is a general negative trend in $\Delta C_\ell$ with increasing $\ell$, as less power has been shifted from large to small scales in the DM$\rightarrow$DR scenario compared to that in $\Lambda$CDM.

We also show in Figure~\ref{fig:deltacell} the induced changes in the linear matter power spectrum $P(k)$, for the same scenario of comparing cosmological parameters.  The suppression of $P(k)$, along with the reduction in $\Omega_m$, can lead to lower values of $S_8$ in the DM$\rightarrow$DR model than in $\Lambda$CDM.  Note that the suppression of $P(k)$ is responsible for the reduced impact of CMB lensing in this model, discussed in the preceding paragraph.

On large scales, we tend to see an increase in $P(k)$. This is due to the modified expansion history in these models: the turnover scale in the power spectrum is set by the scale of the horizon at matter-radiation equality, according to $k_{\mathrm eq}= \frac{a_{\mathrm{eq}}}{a_0}H_{\mathrm{eq}}$, where a subscript $_{\mathrm{eq}}$ refers to quantities at matter-radiation equality. The increase in $H_0$ in models with $\zeta>0$ leads to $k_{\mathrm{eq}}$ (and the low-$k$ matter power spectrum in general) being shifted to larger scales (see also the discussion in~\cite{2016JCAP...08..036P}).

\subsection{Potential to reduce cosmological tensions}

As has been noted previously in the literature, this model has the capacity to simultaneously ameliorate both the $H_0$ and the $S_8$ tensions. First, the $H_0$ tension can be improved because the \textit{Planck} data primarily constrain $\Omega_m h^2$ at recombination but not today. The result of a conversion of some DM to DR after recombination is a lower $\Omega_m$ today (as compared to in $\Lambda$CDM) and thus a slightly higher value of $\Omega_\Lambda$, in order to preserve flatness $\sum_i \Omega_i=1$. This leads to a longer period of exponential expansion due to dark energy, and a higher expansion rate $H_0$ today.  Simultaneously, the decay of some DM leads straightforwardly to less matter clustering today and a lower value of $S_8$.

\begin{figure}[h!]
\includegraphics[width=\columnwidth]{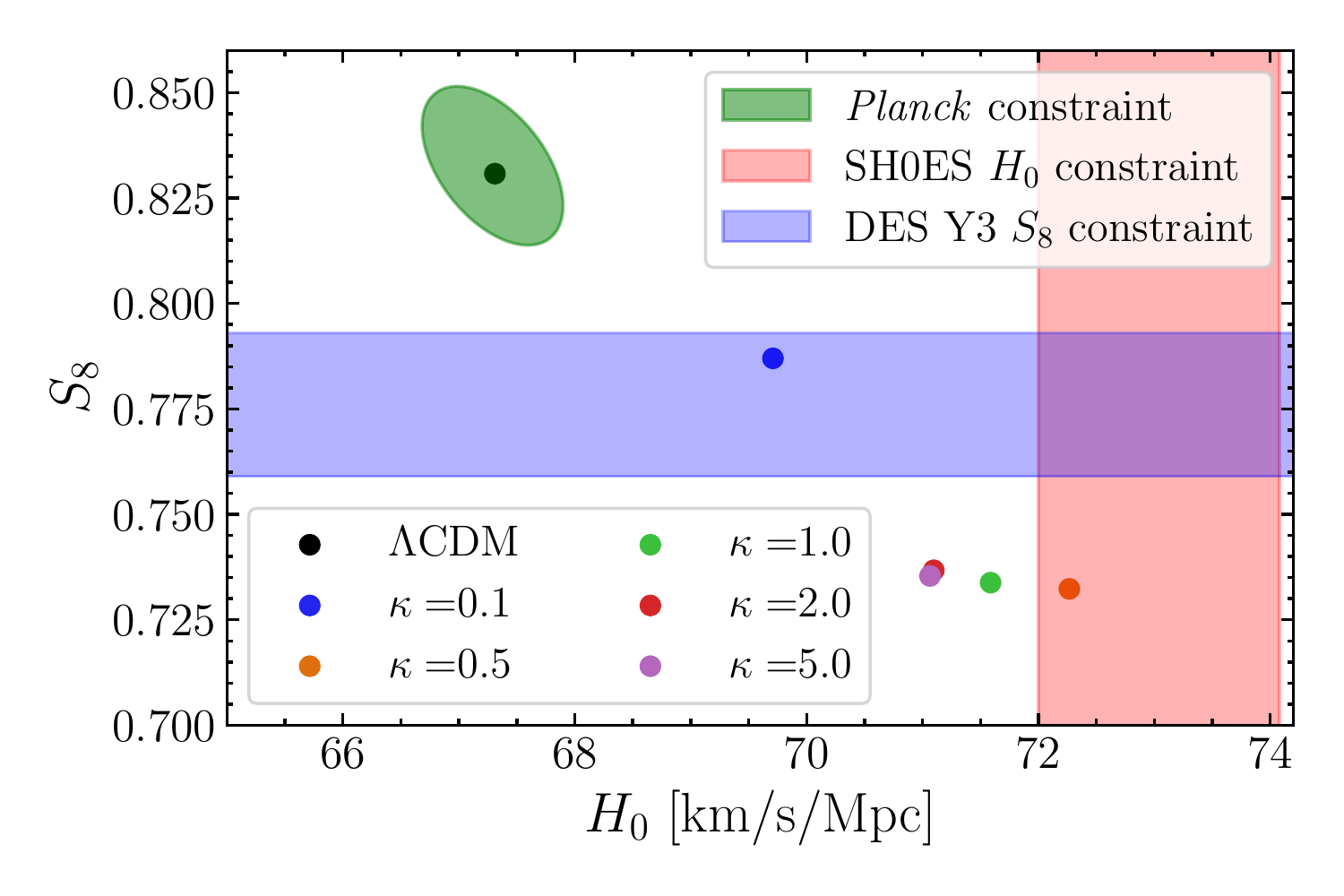}
\caption{The values of $\{H_0,S_8\}$ plotted for the DM$\rightarrow$DR model with $\zeta=0.1$ and  $a_t=0.01$, with the $\Lambda$CDM parameters fixed by the \textit{Planck} 2018 primary CMB analysis. Indicated on the plot is the \textit{Planck} $1\sigma$ error ellipse for $\{H_0,S_8\}$, as well as the SH0ES $H_0$ constraint and the DES-Y3 $S_8$ constraint, which are both in tension with the $\textit{Planck}$ result. We see that the DM$\rightarrow$DR model can change both $H_0$ and $S_8$ in the right direction to reduce these tensions, by increasing $H_0$ and decreasing $S_8$.}\label{fig:S8_H0}
\end{figure}

In Figure~\ref{fig:S8_H0} we show the $\Lambda$CDM $\{H_0,S_8\}$ values (from the \textit{Planck} CMB-only constraints), along with the values of $H_0$ and $S_8$ for the DM$\rightarrow$DR model with $\zeta=0.1$, $a_t=0.01$ and various values of $\kappa$, with the remaining parameters ($\Omega_b h^2$, $\Omega_c^{\star}h^2$, $n_s$, $A_s$) fixed to the \textit{Planck} 2018 primary-CMB values as before (where again by $\Omega_c^\star$ we mean $\Omega_c$ at the time of the release of the CMB). We can see that this model allows for higher values of $H_0$, along with lower values of $S_8$.

\section{Data}\label{sec:data}

We constrain the parameters of the DM$\rightarrow$DR model and compare its performance with $\Lambda$CDM in fitting various cosmological datasets. We include CMB data; large-scale structure (in particular CMB lensing and BAO data); supernovae luminosity distances (which constrain the relative evolution of $H(z)$, i.e., $H(z)/H_0$); a local $H_0$ measurement; and a direct measurement of $S_8$, which is probed by low-$z$ galaxy surveys (or other probes of the matter power spectrum).

In all cases, we calculate the likelihood function with \texttt{CLASS\_DMDR}, with the non-linear matter power spectrum calculated with the Halofit prescription~\cite{2003MNRAS.341.1311S,2012ApJ...761..152T}. This has only a minor impact, as nearly all of the probes we use are dominated by linear modes.

We describe in more detail the datasets and likelihoods we use below.

\subsection{\textit{Planck} primary CMB and CMB lensing}
First, we consider the \textit{Planck} 2018 likelihood for the primary CMB alone~\cite{2020A&A...641A...5P}, including the low-$\ell$ TT, low-$\ell$ EE, and high-$\ell$ TT/TE/EE (plik) power spectra. We also include the 2018 lensing likelihood~\cite{2020A&A...641A...8P} (clik), which probes large-scale structure in the late Universe.

\subsection{Baryon acoustic oscillations}

We use BAO likelihoods from the 6dF galaxy survey~\cite{2011MNRAS.416.3017B} and SDSS DR7~\cite{2015MNRAS.449..835R} and DR12~\cite{2017MNRAS.470.2617A}. The 6dF galaxy survey gives a measurement of the BAO scale at $z=0.106$; SDSS DR7 gives a measurement of the BAO scale at $z=0.15$; DR12 gives measurements at $z=0.38,0.51,0.61$.

\subsection{Luminosity distances: supernovae}

We use supernovae from the Pantheon sample~\cite{2018ApJ...859..101S}, which provide luminosity distances in the redshift range $0.01<z<2.3$.

\subsection{Cosmic distance ladder: $H_0$ from SH0ES}

We use a prior on $H_0$ corresponding to the most recent SH0ES measurement: $H_0=73.04\pm1.04 \,\mathrm{km/s/Mpc}$~\cite{2021arXiv211204510R}, which uses high-$z$ cepheids to calibrate 42 SNe, and 277 SNe in the redshift range $0.023<z<0.15$ to measure $H_0$. The cepheids themselves are calibrated from parallax measurements of cepheids in the Milky Way, the Large and Small Magellanic Clouds, NGC 4258 (geometric megamaser distance), and M31.\footnote{Note that for models in which $H(z)$ is modified at very low redshifts $z \lesssim 0.1$, it is more appropriate to treat the SH0ES likelihood as a prior on the SNe absolute magnitude rather than $H_0$ itself~\cite{2021MNRAS.504.5164C,2021MNRAS.505.3866E}; however, as we do not find strong evidence for the DM$\rightarrow$DR scenario (including any changes to the low-$z$ expansion history), we do not expect this choice to affect our conclusions.}

\subsection{Matter clustering: $S_8$ from DES}

The Dark Energy Survey constrains the low-$z$ Universe with cosmic shear and galaxy clustering data, as well as their cross-correlation (galaxy-galaxy lensing). The DES-Y3 results~\cite{2022PhRvD.105b3520A} combine these to find $S_8= 0.776\pm0.017$. We incorporate this information as a Gaussian prior on $S_8$, in lieu of using the full information in their likelihood. In~\cite{2020A&A...641A...6P} it was explicitly checked (for $\Lambda$CDM and for the early dark energy scenario) that the full DES likelihood can be replaced by the prior on $S_8$ in any analyses that include \emph{Planck}, i.e., the additional cosmological information beyond $S_8$ is negligible compared to that we get from \textit{Planck}.

\section{MCMC analysis and results}\label{sec:results}

 We perform Markov--Chain--Monte--Carlo (MCMC) analyses on the datasets described in Section~\ref{sec:data} using \texttt{Cobaya}~\cite{2019ascl.soft10019T,Torrado:2020dgo}. For the sampling, we use the Metropolis--Hastings algorithm implementation of~\cite{2002PhRvD..66j3511L,2013PhRvD..87j3529L}. We perform our analysis on various subsets of the datasets, and we perform separate $\Lambda$CDM and DM$\rightarrow$DR analyses. In the DM$\rightarrow$DR analyses, we allow all three parameters $\zeta,\kappa,a_t$ to vary.

 We run our chains until they are converged with a Gelman-Rubin convergence criterion~\cite{1992StaSc...7..457G} of $R-1<0.06$.
 
 \subsection*{Priors on the DM$\rightarrow$DR parameters}
 As we are interested in physics that has modified $H(z)$ since recombination, we set a lower bound $a_t > 10^{-4}$ (recall recombination happened at $a\sim10^{-3}$). We sample linearly in $\log_{10} a_t$, and we impose an upper bound $\log_{10}a_t<4$.

   We impose a linear prior on $\zeta$, and allow $0<\zeta<3.16$; this upper bound is sufficient as the data rule out $\zeta<1$ in this regime.
   
    We impose a linear prior on $\kappa$ and let $0<\kappa<4$. While in practice our posteriors do reach this upper bound, we keep it as larger values of $\kappa$ become indistinguishable at the current sensitivity of cosmological data (i.e., the transition is effectively instantaneous). 
    
    We impose the physicality condition $\zeta\le\frac{1}{a_t^\kappa}$ (see Equation~\eqref{physicality_condition}). 
    In principle we could set a larger upper bound on $\log_{10}a_t$, as this would be allowed by the physicality condition at the price of very low $\kappa$. However, as very low $\kappa$ transitions are degenerate with $\Lambda$CDM, this runs the risk of artificially stretching the prior volume around $\Lambda$CDM, which is why we choose a cut-off on $\log_{10}a_t$.

    We use flat, uninformative priors on all standard cosmological and nuisance parameters.
    
    We present the results of our analyses with alternative priors on the DM$\rightarrow$DR parameters in Appendix~\ref{app:prior_effects}; we discuss the effects of the priors at the end of this Section.
   
    \subsection*{Posteriors and results}
\begin{figure*}
 \includegraphics[width=\textwidth]{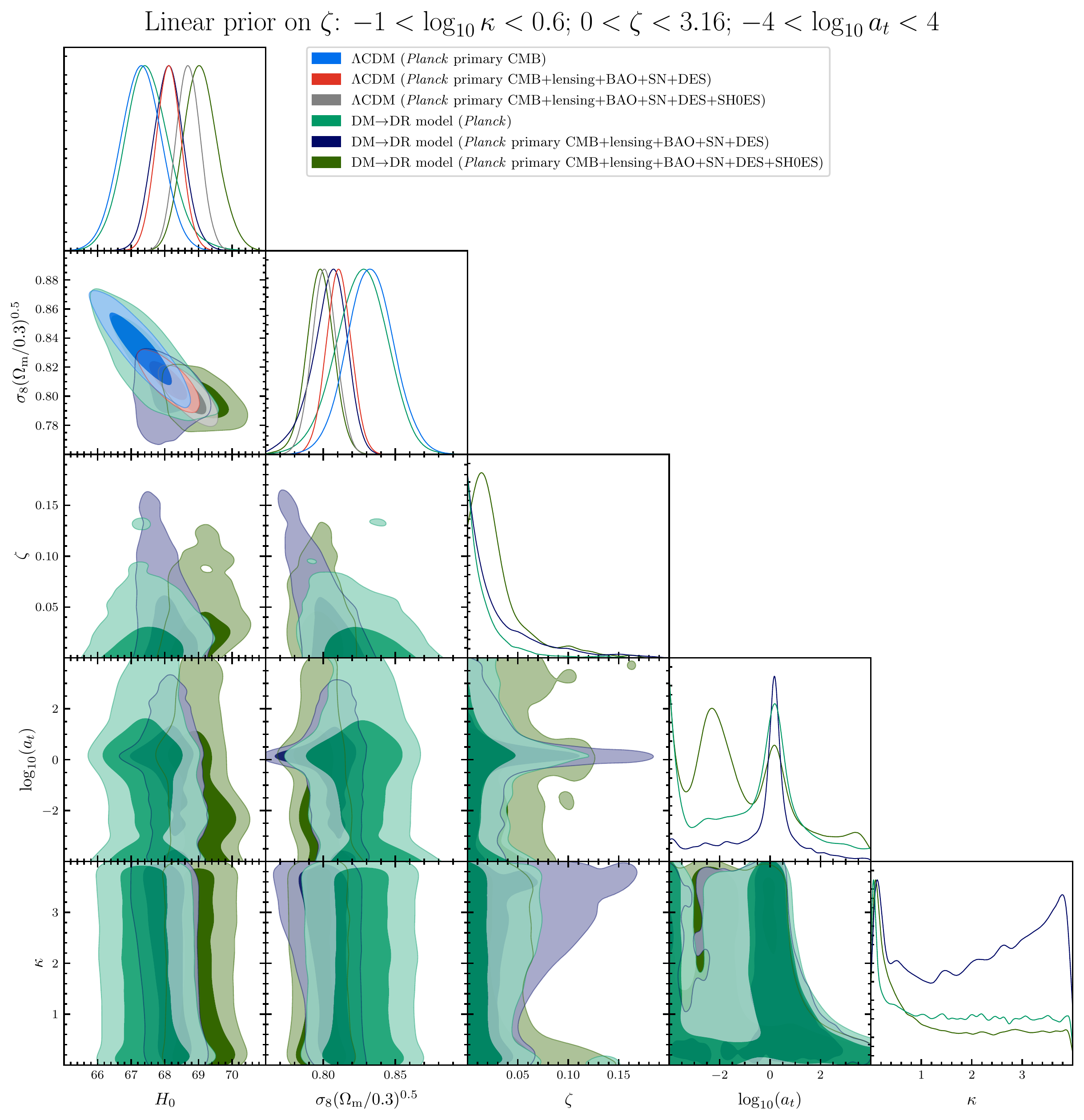}
 \caption{The posteriors on $H_0$ and $S_8$, as well as the DM$\rightarrow$DR parameters, for $\Lambda$CDM analyses and DM$\rightarrow$DR analyses, for our baseline-prior case (linear priors on $\zeta$, $\kappa$, and $\log a_t$, as indicated above the legend). We show the posteriors for the primary-CMB-only (\textit{Planck}) dataset, as well as the combined primary-CMB+lensing+BAO+SH0ES+DES analysis, and the same combination without SH0ES.  }\label{fig:posteriors}
 \end{figure*}

 Our posteriors are shown in Figure~\ref{fig:posteriors}. We see that when we include the SH0ES $H_0$ prior there is a preference for $\zeta>0$, but without this prior the DM$\rightarrow$DR model is \textit{not preferred} over $\Lambda$CDM. Even in the case that there is a preference for $\zeta>0$, the $H_0$ posterior is only slightly shifted with respect to $\Lambda$CDM.  The $S_8$ posteriors are only slightly changed between the $\Lambda$CDM and DM$\rightarrow$DR analyses, with the 1-D posteriors being systematically shifted slightly downwards.  The $68\%$ confidence limits on $H_0$ and $S_8$ shift as indicated in Table~\ref{tab:confidence_limits}.

 \begin{table*}
 \begin{tabular}{|c||c|c||c|c|}\hline
 & \multicolumn{2}{c||}{$H_0$ [km/s/Mpc]}&\multicolumn{2}{c|}{$S_8$}\\\cline{2-5}
 &$\Lambda $CDM& DM$\rightarrow$ DR &$\Lambda $CDM& DM$\rightarrow$ DR\\\hline\hline
\textit{Planck} primary CMB  &  $   66.69<H_0 <67.9   $  & $   66.79<H_0 <68.10   $  & $   0.82<S_8 <0.85   $  & $   0.81<S_8 <0.85   $  \\\hline 
+$\phi\phi$+BAO+SN+DES  &  $   67.74<H_0 <68.49   $  & $   67.65<H_0 <68.55   $  & $   0.80<S_8 <0.82   $  & $   0.79<S_8 <0.82   $  \\\hline 
+SH0ES  &  $   68.33<H_0 <69.06   $  & $   68.53<H_0 <69.53   $  & $   0.79<S_8 <0.81   $  & $   0.79<S_8 <0.81   $  \\\hline 
 
 \end{tabular}
 \caption{$68\%$ confidence limits on $H_0$ and $S_8$, from our baseline analysis.}\label{tab:confidence_limits}
 \end{table*}

 We show in Table~\ref{tab:all_confidence_intervals} the confidence intervals for the base $\Lambda$CDM parameters for the different datasets, along with the DM$\rightarrow$DR parameters and the derived parameters of interest $H_0$, $\Omega_m$, and $S_8$. 
 
  \begin{table*}
\begin{tabular} { |l|  c c |c c| c c|}\hline
&\multicolumn{2}{c|}{\textit{Planck} primary CMB}&\multicolumn{2}{c|}{$+\phi\phi+$BAO+SN+DES}&\multicolumn{2}{c|}{+SH0ES}\\\cline{2-7}
 Parameter &  DM$\rightarrow$DR&$\Lambda$CDM&  DM$\rightarrow$DR&$\Lambda$CDM&  DM$\rightarrow$DR&$\Lambda$CDM\\
\hline\hline

{$\log(10^{10} A_\mathrm{s})$} & $3.046\pm 0.016            $& $3.044\pm 0.016            $& $3.045\pm 0.014            $& $3.041\pm 0.014            $& $3.054^{+0.014}_{-0.015}   $& $3.049^{+0.013}_{-0.015}   $\\

{$n_\mathrm{s}   $} & $0.9641\pm 0.0044          $& $0.9641\pm 0.0044          $& $0.9669\pm 0.0037          $& $0.9679\pm 0.0036          $& $0.9691\pm 0.0044          $& $0.9711\pm 0.0036          $\\

{$\Omega_\mathrm{b} h^2$} & $0.02231\pm 0.00015        $& $0.02234\pm 0.00015        $& $0.02243\pm 0.00014        $& $0.02248\pm 0.00013        $& $0.02247^{+0.00018}_{-0.00016}$& $0.02261\pm 0.00013        $\\

{$\Omega_\mathrm{c} h^2$} & $0.1188^{+0.0026}_{-0.0013}$& $0.1201\pm 0.0014          $& $0.1153^{+0.0042}_{-0.00084}$& $0.11829\pm 0.00082        $& $0.1164^{+0.0017}_{-0.00060}$& $0.11713\pm 0.00079        $\\

{$\tau_\mathrm{reio}$} & $0.0542^{+0.0072}_{-0.0080}$& $0.0541\pm 0.0079          $& $0.0558\pm 0.0072          $& $0.0545\pm 0.0072          $& $0.0601^{+0.0064}_{-0.0079}$& $0.0594^{+0.0068}_{-0.0078}$\\

{$100\theta_\mathrm{s}$} & $1.04185\pm 0.00030        $& $1.04185\pm 0.00030        $& $1.04193\pm 0.00028        $& $1.04198\pm 0.00028        $& $1.04204\pm 0.00030        $& $1.04212\pm 0.00028        $\\

{$\zeta          $} & $< 0.0204                  $&---& $< 0.0374                  $&--- &$< 0.0321                  $&---\\
{$\log_{10}(a_t) $} & $< 0.277                   $&---& $< 0.270                   $& ---&$< -0.0990                 $&---\\
{$\kappa_{dcdm}  $} & \textit{unconstrained}&---                         & \textit{unconstrained}   &---                      & \textit{unconstrained}   &---                   \\\hline\hline

$H_0                       $ & $67.49^{+0.61}_{-0.70}     $& $67.29\pm 0.61             $& $68.11\pm 0.46             $& $68.11\pm 0.38             $& $69.07^{+0.46}_{-0.54}     $& $68.70\pm 0.36             $\\

$\Omega_\mathrm{m}         $ & $0.311^{+0.010}_{-0.0088}  $& $0.316\pm 0.0085          $& $0.2984^{+0.0095}_{-0.0053}$& $0.3049\pm 0.0049          $& $0.2924^{+0.0061}_{-0.0051}$& $0.2975\pm 0.0046          $\\

$\sigma_8 (\Omega_\mathrm{m}/0.3)^{0.5}$ & $0.827\pm 0.018            $& $0.833\pm 0.016            $& $0.803^{+0.014}_{-0.0095}  $& $0.811\pm 0.0088          $& $0.798\pm 0.0096          $& $0.801\pm 0.0085          $\\
\hline\end{tabular}
\caption{The 68\% confidence limits on the base $\Lambda$CDM and DM$\rightarrow$DR parameters, along with the derived parameters $H_0$, $\Omega_m$, and $S_8$, for the DM$\rightarrow$ DR and $\Lambda$CDM analyses.}\label{tab:all_confidence_intervals}
 \end{table*}

\subsection*{$\chi^2$ comparison}\label{sec:chi2comparison}

We maximize the posteriors with the BOBYQA~\cite{bobyqa,2018arXiv180400154C,2018arXiv181211343C} minimizer implemented in \texttt{Cobaya}. The maximum posterior values are listed and compared in Table~\ref{tab:minmaph0s8}. Interestingly, within the accuracy of the minimizer that we use, the minimum of the CMB-only analysis of the DM$\rightarrow$DR model is at $\zeta=0$: i.e., the data fully prefer $\Lambda$CDM over any amount of converting dark matter: the amount of such a component is severely constrained by the CMB. For the extended datasets, this remains true unless the SH0ES prior on $H_0$ is included, in which case the MAP (maximum a posteriori) point is at  $\zeta>0$.

We also list in Table~\ref{tab:minmaph0s8}  the values of the $\chi^2$ of the \textit{Planck} primary CMB alone, at the MAP point of the different analyses. This allows us to directly test whether an extended (\textit{Planck}+LSS) analysis with a modified model could fit the CMB better than a $\Lambda$CDM analysis of a modified model. In particular, a higher value of $H_0$ with a lower $\chi^2_{Planck}$ could indicate that the SH0ES and $\textit{Planck}$ datasets are in less tension within the modified model than within $\Lambda$CDM. However, we find that $\chi^2_{Planck}$ does not improve in the DM$\rightarrow$DR model compared to $\Lambda$CDM.

We also include in Table~\ref{tab:minmaph0s8} the mean values of $H_0$ from the chains, and the values of $H_0$ at the maximum a posteriori (MAP) points. 

\begin{table*}
\begin{tabular}{|c||c|c||c|c||c|c|c|c||c|c|c|c|}
\hline
\multicolumn{1}{|c||}{}&\multicolumn{4}{|c||}{}&\multicolumn{4}{c||}{$H_0$ [km/s/Mpc]}&\multicolumn{4}{c|}{$S_8$}\\\cline{2-13}
&\multicolumn{2}{c||}{$\log(\mathrm{Posterior)}$}&\multicolumn{2}{|c||}{$\chi^2_{Planck}$}&\multicolumn{2}{|c|}{Mean}&\multicolumn{2}{|c||}{MAP}&\multicolumn{2}{|c|}{Mean}&\multicolumn{2}{|c|}{MAP}\\\cline{2-13}
&   $\Lambda$CDM  &   DM$\rightarrow$DR &   $\Lambda$CDM  &   DM$\rightarrow$DR &      $\Lambda$CDM  &   DM$\rightarrow$DR &   $\Lambda$CDM  &   DM$\rightarrow$DR &   $\Lambda$CDM  &   DM$\rightarrow$DR &   $\Lambda$CDM  &   DM$\rightarrow$DR      \\ \hline \hline

\textit{Planck} & 2766.0 & 2766.1 & 2763.8 & 2763.9 & 67.3$\pm$0.6 & 67.5$\pm$0.7 & 67.5 & 67.4 & 0.83$\pm$0.02 & 0.83$\pm$0.02 & 0.83 & 0.83    \\\hline
+LSS & 3821.8 & 3821.7 & 2765.7 & 2765.5 & 68.1$\pm$0.4 & 68.1$\pm$0.5 & 68.1 & 68.0 & 0.81$\pm$0.01 & 0.80$\pm$0.01 & 0.81 & 0.81    \\\hline
+SH0ES & 3841.2 & 3838.4 & 2768.9 & 2769.5 & 68.7$\pm$0.4 & 69.1$\pm$0.5 & 68.7 & 69.4 & 0.80$\pm$0.01 & 0.80$\pm$0.01 & 0.80 & 0.80    \\\hline

\end{tabular}
\caption{The log of the posteriors $\log(\mathrm{Posterior})$ at the maximum a posteriori (MAP) point for the various datasets, along with the $\chi^2$ of the MAP point of the \textit{Planck} primary CMB alone $\chi^2_{Planck}$; note that the calculation contains contributions from the nuisance parameters, so even for the CMB-alone analysis $\chi^2_{Planck}$ does not exactly coincide with $\log(\mathrm{Posterior})$. We also show the mean value of $H_0$ from the MCMC chains, along with its value at the MAP; similarly for $S_8$. The error bars indicated on the mean are calculated from the covariance of the MCMC chains. Note that, in order to save space, we have used \textit{Planck} to refer to the \textit{Planck} primary CMB likelihood, and $+$LSS to denote $+\phi\phi$+BAO+SN+DES. }\label{tab:minmaph0s8}
\end{table*}

We show in Table~\ref{tab:allbestfits_lcdm} the values of the base $\Lambda$CDM and DM$\rightarrow$DR parameters, along with the derived parameters of interest $H_0$, $\Omega_m$, and $S_8$. It should be noted that, within the accuracy of the minimizer we use, the MAP of the DM$\rightarrow$DR case finds $\zeta=0$ exactly for the primary-CMB-alone and the CMB+LSS analyses; i.e., the best-fit points are exactly at $\Lambda$CDM.

We show in Table~\ref{tab:table_dchi2} the $\chi^2$ values at the MAP of the full analysis with all datasets for both $\Lambda$CDM and DM$\rightarrow$DR. We see that the DM$\rightarrow$DR model does have the capacity to reduce $\chi^2_{H_0}$, from 17.5 to 12.2, with even a slight decrease in the \textit{Planck} high-$\ell$ $\chi^2$; however, we do see a slight increase in the low-$\ell$ $\chi^2$ (due to the increased ISW effect in DM$\rightarrow$DR) and in the BAO $\chi^2$ (due to the lower matter density and higher DE density).

\begin{table*}
\begin{tabular}{ |l|  c c |c c| c c|}\hline
&\multicolumn{2}{c|}{\textit{Planck} primary CMB}&\multicolumn{2}{c|}{$+\phi\phi+$BAO+SN+DES}&\multicolumn{2}{c|}{+SH0ES}\\\cline{2-7}
 Parameter &  DM$\rightarrow$DR&$\Lambda$CDM&  DM$\rightarrow$DR&$\Lambda$CDM&  DM$\rightarrow$DR&$\Lambda$CDM\\\hline\hline

$\log(10^{10} A_\mathrm{s})$   &   3.03944   &   3.04409   &   3.0405   &   3.03819   &   3.05271   &   3.04322\\
$n_\mathrm{s}   $   &   0.96551   &   0.96587   &   0.96807   &   0.96771   &   0.96432   &   0.97283\\
$\Omega_\mathrm{b} h^2$   &   0.02236   &   0.02236   &   0.02247   &   0.02247   &   0.02235   &   0.02262\\
$\Omega_\mathrm{c} h^2$   &   0.11973   &   0.11967   &   0.11846   &   0.11828   &   0.11761   &   0.11712\\
$\tau_\mathrm{reio}$   &   0.05414   &   0.05525   &   0.05352   &   0.05347   &   0.05791   &   0.05569\\
$100\theta_\mathrm{s}$   &   1.04185   &   1.04184   &   1.04192   &   1.04197   &   1.04185   &   1.0421\\
$\zeta$   &   0.0   & ---     &   0.0   & ---     &   0.01986   & ---  \\
$\log_{10}(a_t) $   &   /   & ---     &   /   & ---     &   -2.49691   & ---  \\
$\kappa$   &   /   & ---     &   /   & ---     &   3.97581   & ---  \\\hline\hline
$\Omega_\mathrm{m}$   &   0.314   &   0.313   &   0.306   &   0.305   &   0.292   &   0.297\\
$H_0$   &   67.45   &   67.46   &   68.02   &   68.10   &   69.41   &   68.70\\
$S_8$   &   0.826   &   0.828   &   0.813   &   0.810   &   0.799   &   0.799\\\hline

\end{tabular}
\caption{The values of the base $\Lambda$CDM and DM$\rightarrow$DR parameters, along with the derived parameters $H_0$, $\Omega_m$, and $S_8$, at the MAP points for each analysis. The minimizer finds the $\Lambda$CDM point ($\zeta=0$) for the cases with no $H_0$ prior from SH0ES, although note that there are small (sub-percent, except for the case of the least-constrained parameter $\tau$) residual differences in the location of the best-fit point due to the numerical tolerance of our minimizer.  }\label{tab:allbestfits_lcdm}
\end{table*}

\begin{table}
\begin{tabular}{|c|c|c|}\hline
Likelihood & $\Lambda$CDM &DM$\rightarrow$DR\\\hline\hline
\textit{Planck} TT+TE+EE (high $\ell$)      &      2350.9      &      2348.4     \\\hline
\textit{Planck} low-$\ell$ TT      &      22.0      &      24.3     \\\hline
\textit{Planck} low-$\ell$ EE      &      396.0      &      396.8     \\\hline\hline
\textbf{Total primary CMB}   &  2768.9      &      2769.4\\\hline\hline
\textit{Planck} $\phi\phi$      &      10.2      &      10.4    \\\hline\hline
SDSS BAO:  6dF      &      0.0340      &      0.1134     \\\hline
SDSS BAO:  DR7 MGS      &      2.32      &      2.84     \\\hline
SDSS BAO: DR12 consensus      &      3.53      &      4.34     \\\hline\hline
\textbf{Total BAO}   &  5.88      &      7.29\\\hline\hline
Pantheon SNIa      &      1034.7      &      1034.8     \\\hline\hline
DES-Y3: $S_8$      &      1.795      &      1.826     \\\hline\hline
SH0ES: $H_0$      &      17.51      &      12.22     \\\hline\hline
\textbf{Total}   &  3839.1      &      3836.1\\\hline
\end{tabular}
\caption{The $\chi^2$ values of the different likelihoods at the MAP points for the joint analysis of all datasets (i.e., the parameters shown in the last column of Table~\ref{tab:allbestfits_lcdm}), for both $\Lambda$CDM and DM$\rightarrow$DR.  }\label{tab:table_dchi2}
\end{table}

\subsection*{Prior dependence}

In principle, the posteriors of our analysis depend on the choice of prior. We must choose a prior on $\zeta$, $a_t$, and $\kappa$. As the values of $a_t$ we are interested in exploring span several orders of magnitude, it is natural to choose a linear prior on $\log_{10}a_t$. $\zeta$ and $\kappa$ are more complicated: we expect $\zeta$ to be small and so a linear prior on $\log_{10}\zeta$ might at first appear natural; however, small values of $\zeta$ are degenerate with $\Lambda$CDM and indistinguishable within the data. Indeed, the data are constraining enough to prefer small values of $\zeta$ regardless of the prior, and we find that a linear prior on $\zeta$ is perhaps more appropriate as it avoids the prior-volume-stretching effects of the logarithmic prior.

\begin{figure*}
 
 \includegraphics[width=\columnwidth]{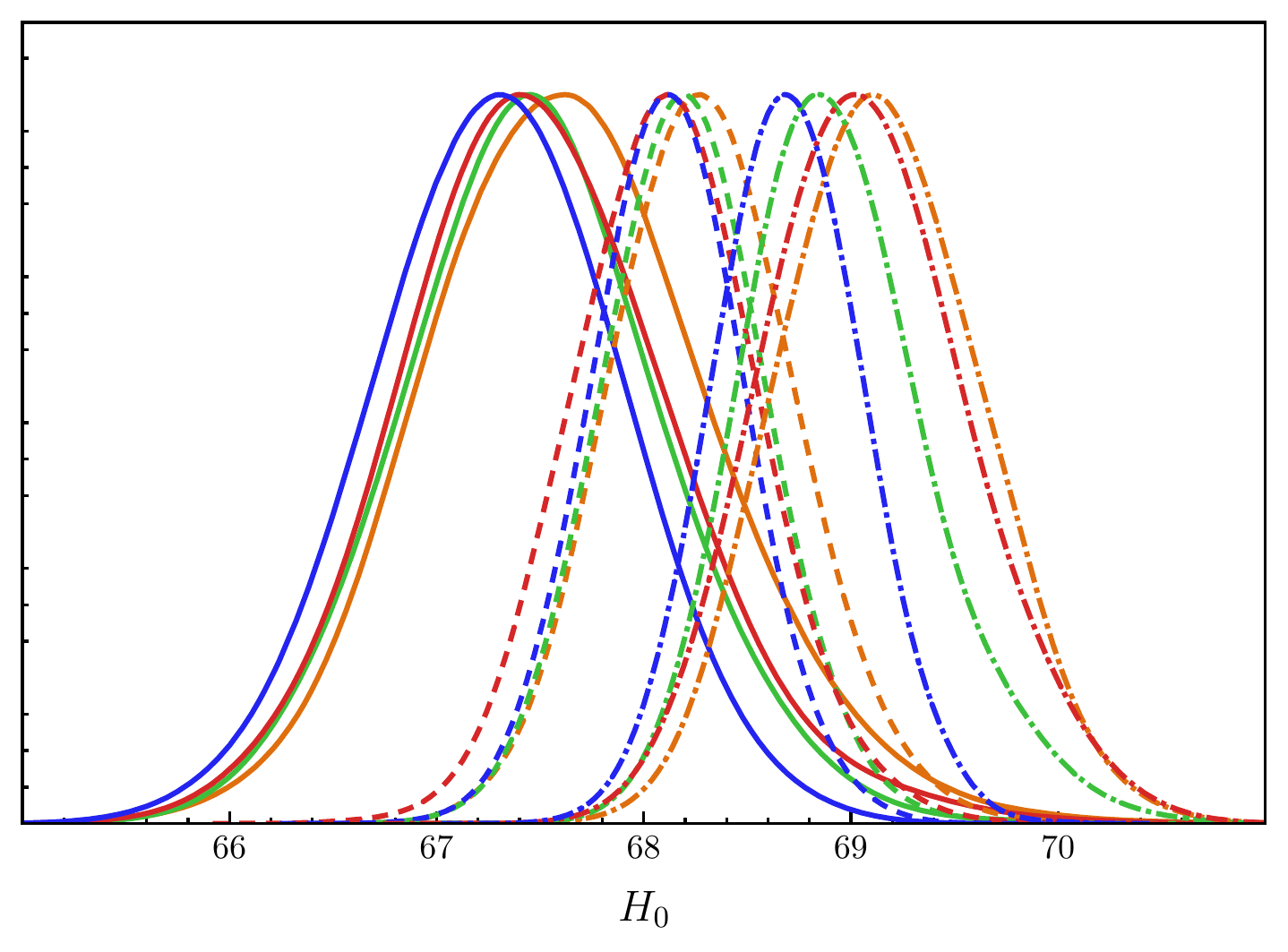}
 \includegraphics[width=\columnwidth]{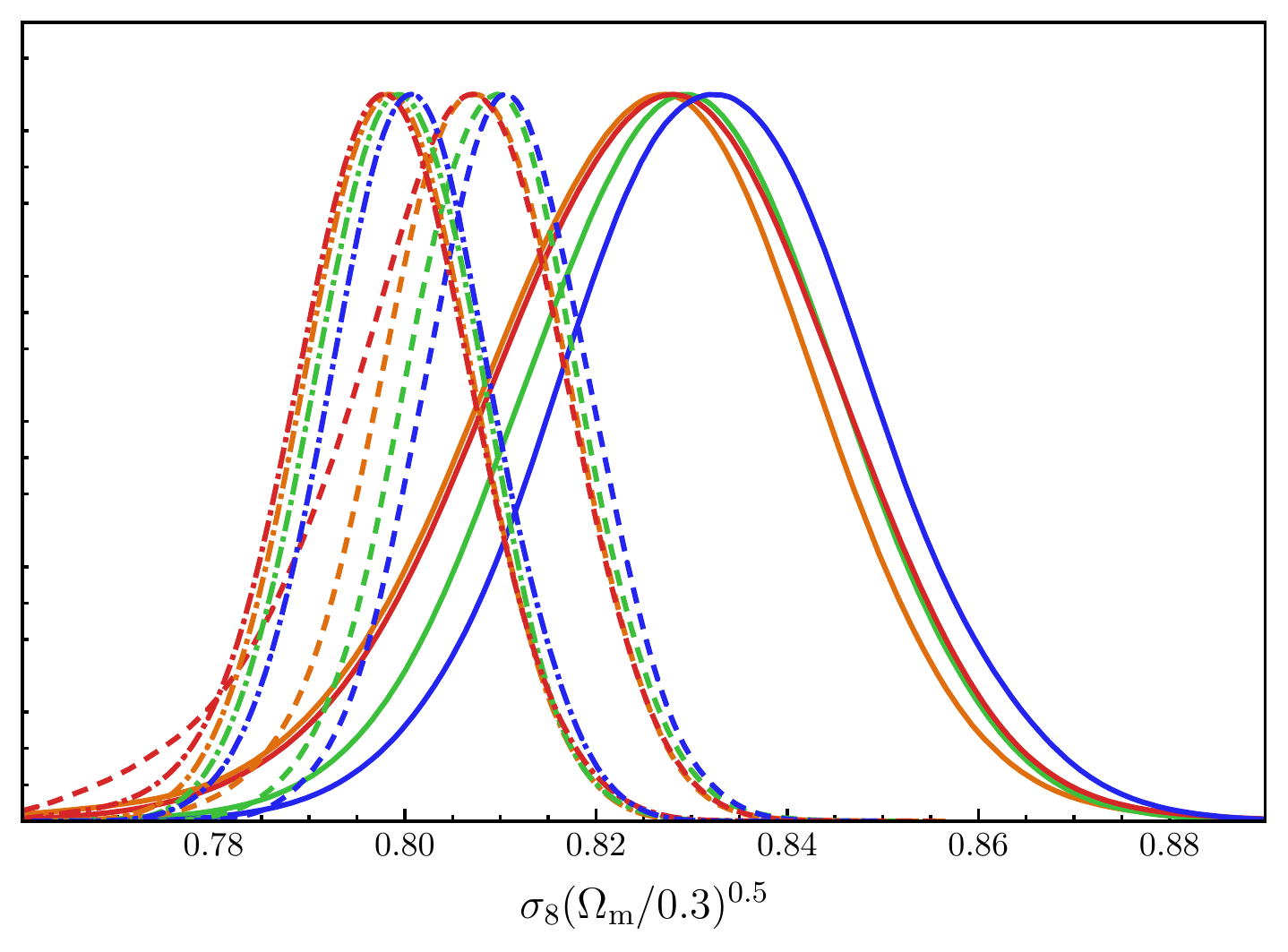}\\
 \includegraphics[width=\columnwidth]{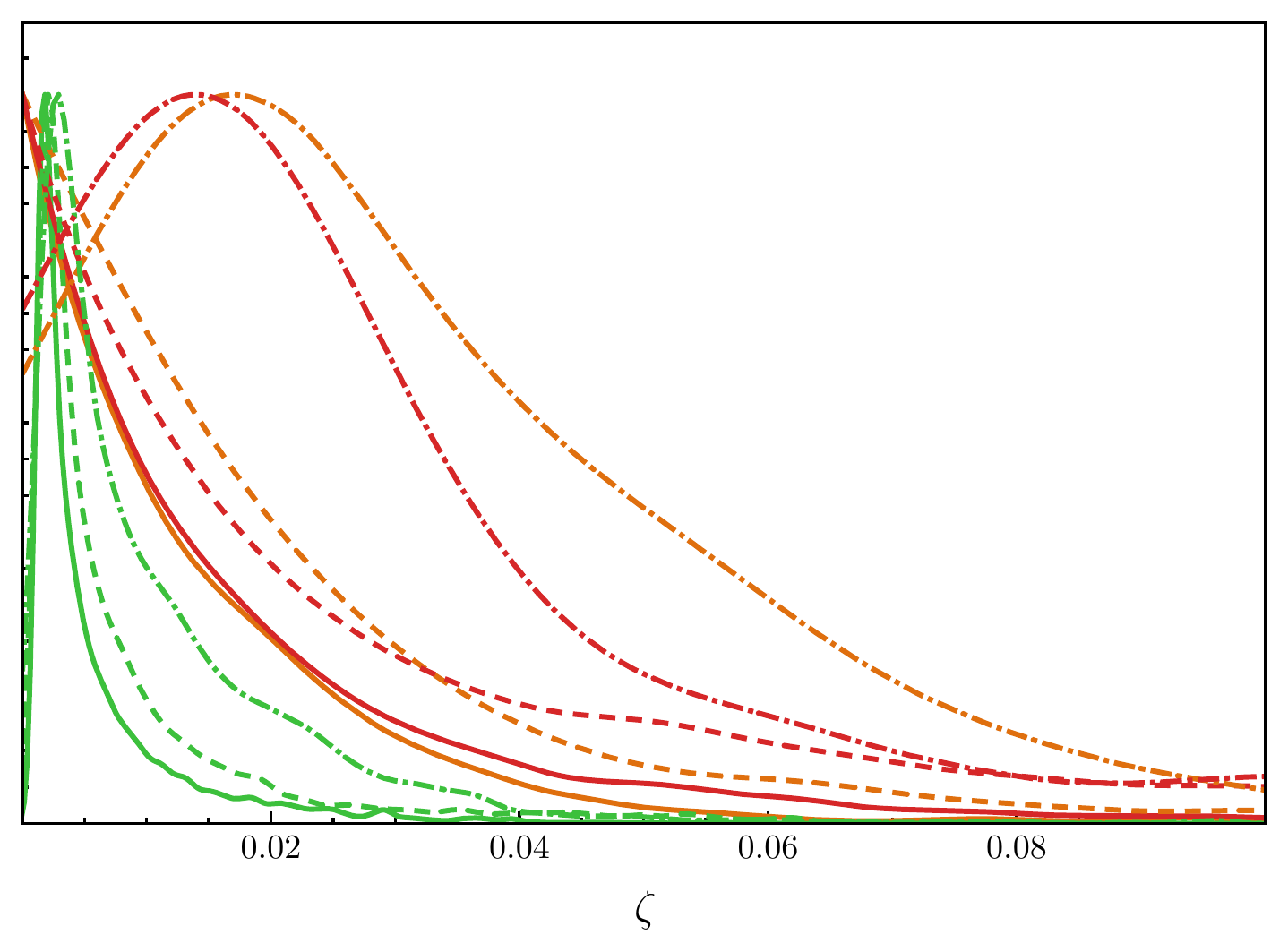}
 \includegraphics[width=\columnwidth]{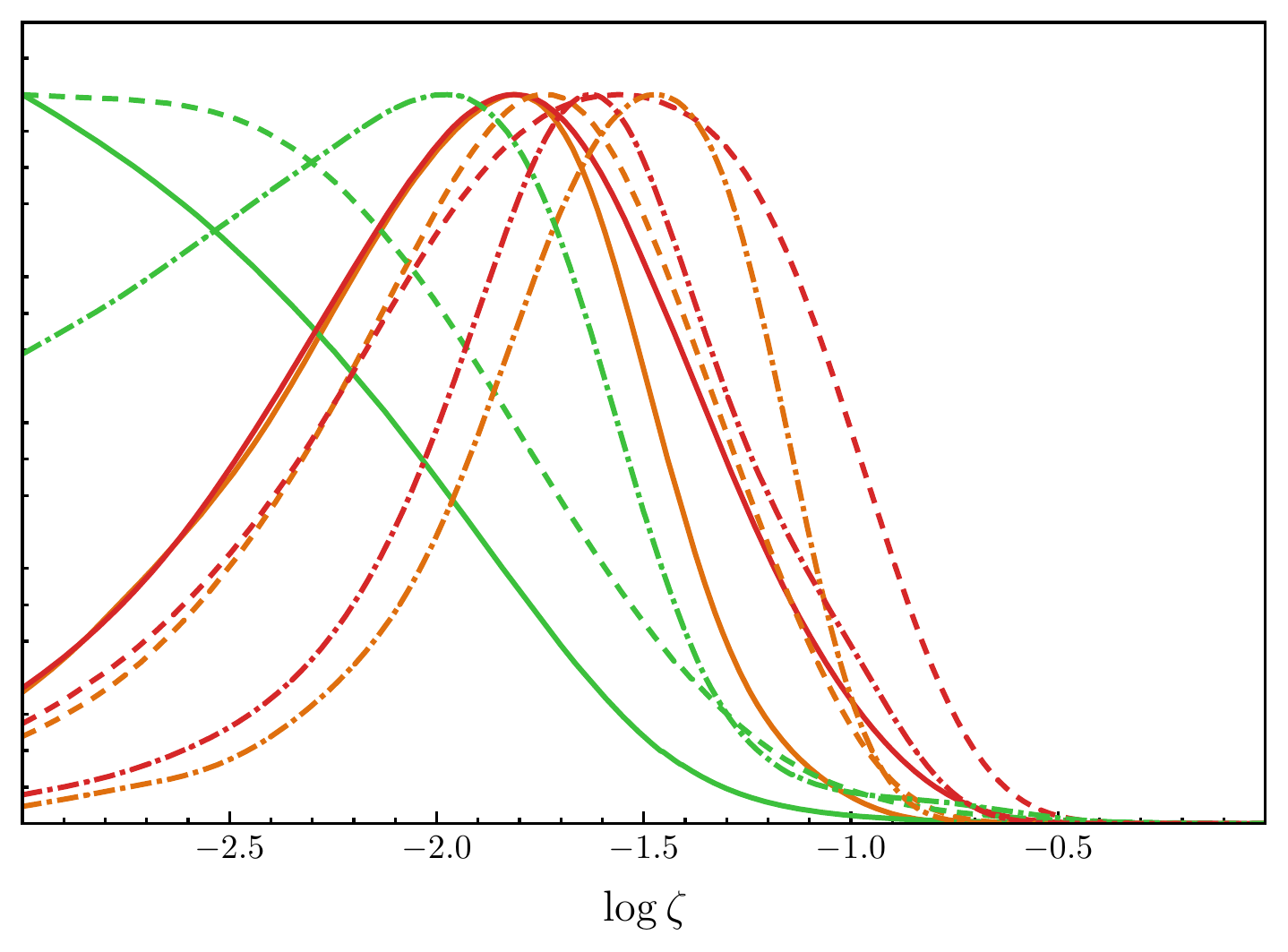}
 \includegraphics[width=0.5\columnwidth]{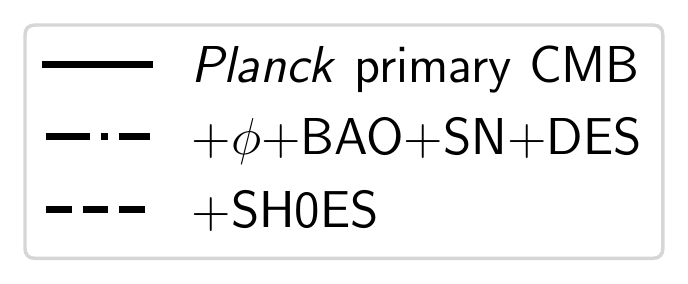}
 \includegraphics[width=0.5\columnwidth]{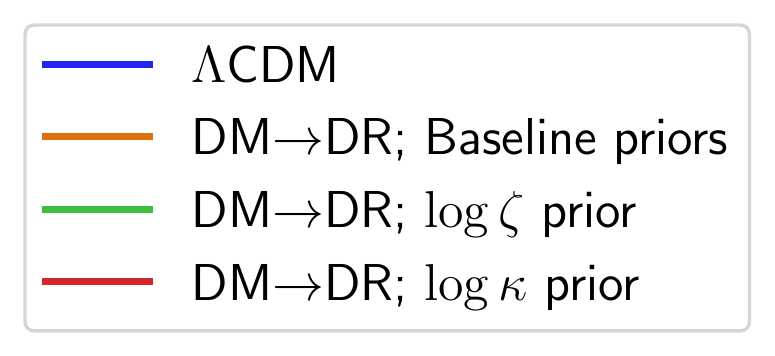}
 \caption{The effect of prior choice on the 1-D $H_0$ (top left); $S_8$ (top right); $\zeta$ (bottom left); and $\log\zeta$ (bottom right) posteriors. Our main conclusion is that the $\log\zeta$-prior analyses (in green) prefer lower $\zeta$ than the linear-$\zeta$-prior analyses.  }\label{fig:prior_comparison}
\end{figure*}

To check for prior dependence, we have performed our analysis on  three choices of prior (linear in $\zeta$ and $\kappa$; linear in $\log_{10}\zeta$ and $\kappa$; and linear in $\zeta$ and $\log_{10}\kappa$). While we choose the first option (linear in $\zeta$ and $\kappa$) for our baseline results included in this Section, we present the results for the alternative choices in Appendix~\ref{app:prior_effects}. Summaries of the effects on various parameters of interest ($H_0$; $S_8$; and $\zeta$) are shown in Figure~\ref{fig:prior_comparison}, by explicitly plotting the 1-D
posteriors of these parameters. A general conclusion is that the linear prior on $\log_{10}\zeta$ causes the sampler to spend more time in the $\Lambda$CDM region and thus prefers lower values of $\log_{10}\zeta$ that are not reached by the sampler that is linear in $\zeta$; this is clear from the bottom right plot, the posterior of $ \log_{10}\zeta$. This results in the $H_0$ and $S_8$ posteriors being slightly less shifted with respect to the $\Lambda$CDM posteriors in the logarithmic-$\zeta$-prior versus linear-$\zeta$-prior cases, as evident in the upper two plots where the  posteriors corresponding to the $\log\zeta$ prior (in green) are in all cases in between the $\Lambda$CDM (blue) and baseline linear prior (orange) posteriors. There is less difference between the logarithmic-$\kappa$ prior case and the linear-$\kappa$ (baseline) case; the $\zeta$ posteriors are slightly lower, possibly due to the preferential exploration lower values of $\kappa$ resulting in the higher values of $a_t$ (and those lower values of $\zeta$) allowed by the physicality condition.

 \subsection*{Failure of the model}
 
\begin{figure}[h!]
     \includegraphics[width=\columnwidth]{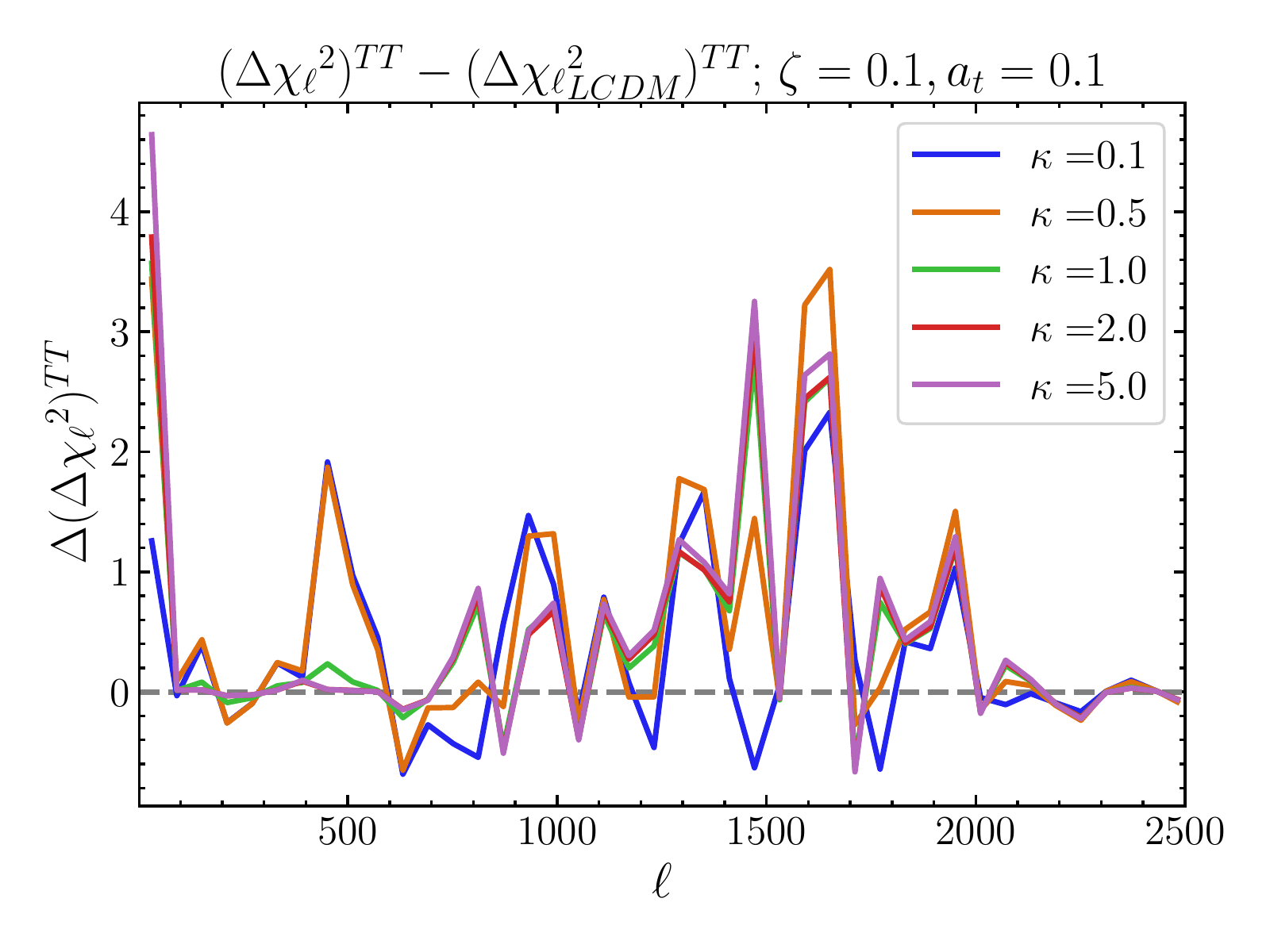}
     \caption{An indicator of the relative $\Delta\chi^2_{TT}$ as a function of $\ell$ between the DM$\rightarrow$DR power spectra shown in Figure~\ref{fig:deltacell} and the \textit{Planck} $TT$ data. We sum the $\Delta\chi^2$ values in $\ell$-bins of width $\Delta\ell=60$. In each case we subtract the equivalent $\Lambda$CDM quantity from the DM$\rightarrow$DR quantity and divide by the $\Lambda$CDM quantity. As such, positive values on this plot indicate a worse fit than $\Lambda$CDM and negative values a better fit. It is clear that the bulk of the ``badness of fit'' of DM$\rightarrow$DR is sourced by the intermediate-to-high-$\ell$ regime where CMB lensing is changing the power spectra; the ISW effect also causes a spike at low $\ell$.  }\label{fig:deltachi2_comparison}
 \end{figure}

 The DM$\rightarrow$DR model, along with various other models in which CDM converts to DR in the late Universe, are severely constrained by the primary CMB data from \textit{Planck}. This is because the primary CMB power spectrum is not solely a probe of the early Universe, but carries late-Universe information through the ISW effect and the effects of CMB lensing on the power spectrum. We show in Figure~\ref{fig:deltachi2_comparison} the relative ``badness-of-fit'' as a function of $\ell$ for the same parameter configurations we have been considering in our previous plots. In this plot, we have calculated the mean $\Delta \chi_\ell^2$ in $\ell$ bins of width $\Delta\ell=60$, for both $\Lambda$CDM and DM$\rightarrow$DR, with respect to the \textit{Planck} measurements of the primary CMB temperature power spectrum. As in Figure~\ref{fig:deltacell}, we are plotting quantities relative to $\Lambda$CDM; positive values on this plot are indicative of a worse fit to the \textit{Planck} data than $\Lambda$CDM, and negative values would indicate a better fit. There is clearly a larger $\Delta \chi^2$ in DM$\rightarrow$DR than $\Lambda$CDM at low $\ell$ (caused by the excess ISW effect) and at medium-to-high $\ell$ ($1000 \lesssim \ell \lesssim 2000$). Indeed, most of the contribution to the higher $\Delta\chi^2$ in DM$\rightarrow$DR comes from this medium-to-high-$\ell$ regime, where the physical effect causing the difference between the spectra is the amount of CMB lensing. In general, the CMB temperature power spectrum is so tightly constrained by \textit{Planck} that it is hard to change any physics affecting structure growth at $z \greaterthanapprox 1$ where CMB lensing is the most efficient (this is why, without a $H_0$ prior, the $\log a_t$ posteriors in Figure~\ref{fig:posteriors_logkappa} tend to peak around $a_t=1$, i.e., today), meaning that the data prevent this model from differing enough from $\Lambda$CDM to meaningfully change $H_0$.
 
\section{Conclusions / Discussion}\label{sec:conclusion}

Late-Universe decays of dark matter have been suggested in the literature as possible paths to cosmological concordance~\cite{2019PhRvD..99l1302V}. Previous studies that have focused on using CMB data to constrain directly the amount of decaying dark matter have placed stringent bounds on the amount of such a component in our Universe~\cite{2014JCAP...12..028A,2016JCAP...08..036P}, and have found that no more than a few percent of the CDM that was present at recombination could have decayed before today. In general, investigations of whether such a scenario can relieve the well-known cosmological tensions have found that they cannot~\cite{2021PhRvD.103d3014C,2021PhRvD.104l3533A,2021JCAP...05..017N}.

In our work, we have performed a very general analysis of a scenario in which some dark matter has converted into dark radiation after the release of the CMB, in a model-agnostic parametrization. Our results agree with studies of the more specific decaying dark matter case, although they are more general: we find that no model in which some component of the CDM present at recombination converts into DR before today will solve these tensions. 
In our work, we have extended and generalized other investigations of such a model~\cite{2022arXiv220304818A}, and corrected some aspects of previous implementations of this scenario.  

While it is true that indeed such a scenario can provide the correct background evolution to reduce these tensions, it is the perturbations that rule them out, in particular tight constraints on the medium-to-high-$\ell$ CMB power spectrum where lensing is important, along with the low-$\ell$ ISW contributions to the CMB perturbations. To avoid  these late-Universe contributions, one could modify our scenario to one in which the CDM converts into a warm DM component, i.e., a DM particle with a small but non-zero mass. However, in avoiding the extra ISW and reduced CMB lensing contributions we would also avoid the attractive aspects of this model, from a concordance standpoint: the Universe would remain in matter domination longer, and not increase $H_0$, although it is true that the $S_8$ tension could still be mitigated by the free-streaming of the warm DM particles~\cite{2020arXiv200809615A}.

\begin{acknowledgements}
We thank Angela Chen, Dragan Huterer, Julien Lesgourgues, and Meng-Xiang Lin for useful conversations.  JCH acknowledges support from NSF grant AST-2108536, NASA grant 21-ATP21-0129, the Sloan Foundation, and the Simons Foundation.  We thank the Scientific Computing Core staff at the Flatiron Institute for computational support.  The Flatiron Institute is supported by the Simons Foundation.
\end{acknowledgements}

\appendix

\section{Calculation of the hypergeometric function}\label{app:hypergeometric}

The DR density in our scenario is
\begin{align}
&\rho_{DR}(a) = \zeta\frac{\rho_{DM}^0}{a^3}\frac{\lb 1+a_t^\kappa\rb}{\lb a^\kappa+a_t^\kappa\rb}\times\nonumber\\
&\,\,\left(\lb a^\kappa+a_t^\kappa\rb {}_2F_1\left[1,\frac{1}{\kappa};1+\frac{1}{\kappa};-\lb\frac{a}{a_t}\rb^\kappa\right]-a_t^\kappa\right)\,.
\end{align}
Calculating this in \texttt{Class\_DMDR} requires the evaluation of the hypergeometric function
\begin{equation}
_2F_1\left[1,\frac{1}{\kappa};1+\frac{1}{\kappa};-\lb\frac{a}{a_t}\rb^\kappa\right].\label{hyperf21}
\end{equation}
To calculate this, we use the GSL function \texttt{gsl\_sf\_hyperg}~\cite{gough2009gnu}. In this Appendix we discuss some subtleties of the implementation.

As \texttt{gsl\_sf\_hyperg} is only defined for arguments $x$ of the hypergeometric function $_2F_1(a,b;c;x)$ with $|x|<1$, in the regime that $a>a_t$ we use the identity
\begin{equation}
_{2}F_1(a,b;c;x)=(1-x)^{-a}{}_2F_1\left(a,c-b;c;\frac{x}{x-1}\right)
\end{equation}
to write expression~\eqref{hyperf21} as
\begin{equation}
\frac{1}{1+\left(\frac{a}{a_t}\right)^\kappa}{}_2 F_1\left[1,1;1+\frac{1}{\kappa};\frac{\left(\frac{a}{a_t}\right)^\kappa}{\left(\frac{a}{a_t}\right)^\kappa+1}\right].\label{2f1xgreaterthan1}
\end{equation}
The numerical calculation of the hypergeometric function $_2F_1(a,b;c;x)$ is inefficient  when $|x|$ is close to 1; to avoid these inefficiencies, where $\left(\frac{a}{a_t}\right)^\kappa>100$ we use the identity
\begin{widetext}
\begin{align}
_2 F_1(a,b;c;x) =\frac{\Gamma(c)\Gamma(c-a-b)}{\Gamma(c-a)\Gamma(c-b)}{}&_2F_1\left(a,b;a+b+1-c;1-x\right)\nonumber\\&+\frac{\Gamma(c)\Gamma(a+b-c)}{\Gamma(a)\Gamma(b)}(1-x)^{c-a-b}{}_2F_1\left(c-a,c-b;1+c-a-b;1-x\right),
\end{align}
\end{widetext}
where $\Gamma(x)$ is the gamma function, to write expression~\eqref{2f1xgreaterthan1} as
\begin{widetext}
\begin{align}
\left(\frac{1}{1+\left(\frac{a}{a_t}\right)^\kappa}\right)&\bigg{(}\frac{\Gamma\left(1+\frac{1}{\kappa}\right)\Gamma\left(-1+\frac{1}{\kappa}\right)}{\Gamma\left(\frac{1}{\kappa}\right)\Gamma\left(\frac{1}{\kappa}\right)}{}_2F_1\left[1,1;2-\frac{1}{\kappa};\frac{1}{\left(\frac{a}{a_t}\right)^\kappa+1}\right]\nonumber\\
&+\Gamma\left(1+\frac{1}{\kappa}\right)\Gamma\left(1-\frac{1}{\kappa}\right)\left(\frac{1}{1+\left(\frac{a}{a_t}\right)^\kappa}\right)^{\frac{1}{\kappa}-1}{}_2F_1\left[\frac{1}{\kappa},\frac{1}{\kappa};\frac{1}{\kappa};\frac{1}{\left(\frac{a}{a_t}\right)^\kappa+1}\right]\bigg{)}.
\end{align}
\end{widetext}
We then approximate the hypergeometric functions as 1 (as $_2F_1(a,b,c,0)=1$) and $1+\left(\frac{a}{a_t}\right)^\kappa$ as $\left(\frac{a}{a_t}\right)^\kappa$ and write this as 
\begin{widetext}
\begin{equation}
    \left(\frac{1}{\left(\frac{a}{a_t}\right)^\kappa}\right)\frac{\Gamma\left(1+\frac{1}{\kappa}\right)\Gamma\left(-1+\frac{1}{\kappa}\right)}{\Gamma\left(\frac{1}{\kappa}\right)\Gamma\left(\frac{1}{\kappa}\right)}+\left(\frac{1}{\left(\frac{a}{a_t}\right)}\right)\Gamma\left(1+\frac{1}{\kappa}\right)\Gamma\left(1-\frac{1}{\kappa}\right).
\end{equation}
\end{widetext}
We use the GSL function \texttt{gsl\_sf\_gamma} to calculate the gamma functions.

\section{Priors on the DM$\rightarrow$DR parameters}\label{app:prior_effects}

In this Appendix we explore the effects of the prior chosen on the DM$\rightarrow$DR parameters on the analysis. In particular, we explore whether using a linear prior on $\zeta$ or $\log_{10}\zeta$ changes the results of the analysis, and similarly whether a linear prior on $\kappa$ or $\log_{10}\kappa$ changes the results of the analysis.

\subsection{Linear prior on $\log\zeta$}

\begin{figure*}
\includegraphics[width=\textwidth]{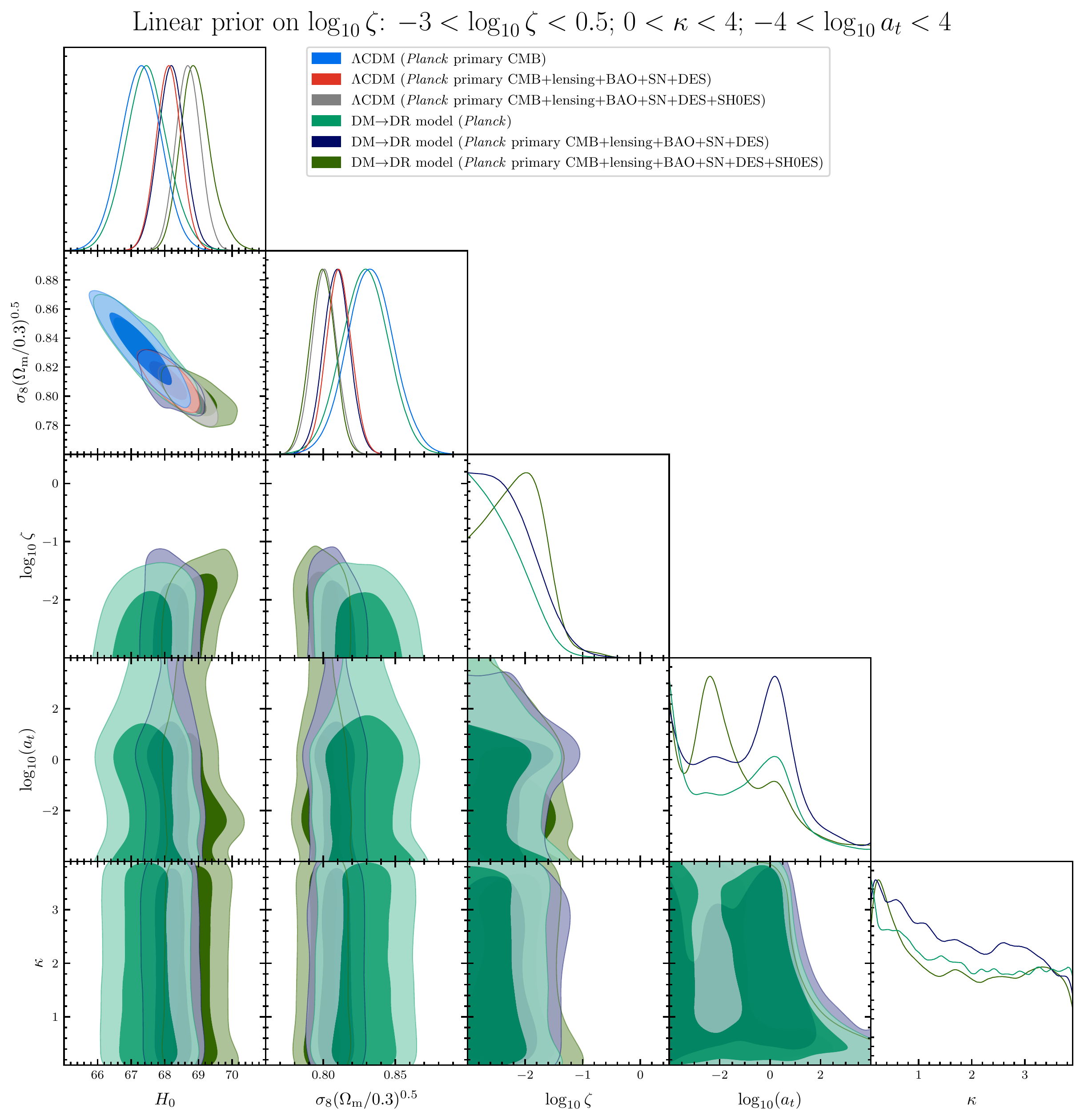}
\caption{The posteriors on $H_0$ and $S_8$, as well as the DM$\rightarrow$DR parameters, for $\Lambda$CDM analyses and DM$\rightarrow$DR analyses, in the case when we use a linear prior on $\log_{10}\zeta$ (cf.~our baseline case in Figure~\ref{fig:posteriors} where we use a linear prior on $\zeta$).}\label{fig:posteriors_logzeta}
\end{figure*}

In Figure~\ref{fig:posteriors_logzeta} we use a linear prior on $\log_{10}\zeta$, while in the main text (and for our main results) we used a linear prior on $\zeta$. We take a lower bound on the prior of $\log_{10}\zeta>-3$, and an upper bound of $\log_{10}\zeta=0.5$ (corresponding to our linear prior of $0<\zeta<3.16$). While we could in principle choose a lower bound on the prior than $\log_{10}\zeta>-3$, we impose this prior to avoid artificially stretching the parameter volume as low values of $\zeta$ are degenerate with each other (and $\Lambda$CDM). We retain the linear priors on $\kappa$ and $\log_{10}a_t$ used in our main analysis: $0<\kappa<4$ and $-4<\log_{10}a_t<4$, along with the physicality condition $\zeta<\frac{1}{a_t^\kappa}$.

We present the 68\% confidence intervals on $H_0$ and $S_8$ in Table~\ref{tab:confidence_limits_logzeta} (in analogy to Table~\ref{tab:confidence_limits} for the baseline linear prior on $\zeta$). We present the $68\%$ confidence intervals on the 6 base $\Lambda$CDM parameters and the DM$\rightarrow$DR parameters in Table~\ref{tab:all_confidence_intervals_logzeta} (in analogy to Table~\ref{tab:all_confidence_intervals}).

\begin{table*}
 \begin{tabular}{|c||c|c||c|c|}\hline
 & \multicolumn{2}{c||}{$H_0$ [km/s/Mpc]}&\multicolumn{2}{c|}{$S_8$}\\\cline{2-5}
 &$\Lambda$CDM& DM$\rightarrow$ DR &$\Lambda$CDM& DM$\rightarrow$ DR\\\hline\hline
\textit{Planck} primary CMB  &  $   66.69<H_0 <67.9   $  & $   66.82<H_0 <68.07   $  & $   0.82<S_8 <0.85   $  & $   0.81<S_8 <0.85   $  \\\hline 
+$\phi\phi$+BAO+SN+DES  &  $   67.74<H_0 <68.49   $  & $   67.79<H_0 <68.59   $  & $   0.80<S_8 <0.82   $  & $   0.80<S_8 <0.82   $  \\\hline 
+SH0ES  &  $   68.33<H_0 <69.06   $  & $   68.43<H_0 <69.31   $  & $   0.79<S_8 <0.81   $  & $   0.79<S_8 <0.81   $  \\\hline 
 \end{tabular}
 \caption{$68\%$ confidence limits on $H_0$ and $S_8$, from the analysis with a linear prior on $\log\zeta$; cf.~Table~\ref{tab:confidence_limits} for the analogous results for the baseline analysis with a linear prior on $\zeta$.}\label{tab:confidence_limits_logzeta}
 \end{table*}

  \begin{table*}
\begin{tabular} { |l|  c c |c c| c c|}\hline
&\multicolumn{2}{c|}{\textit{Planck} primary CMB}&\multicolumn{2}{c|}{$+\phi\phi+$BAO+SN+DES}&\multicolumn{2}{c|}{+SH0ES}\\\cline{2-7}
 Parameter &  DM$\rightarrow$DR&$\Lambda$CDM&  DM$\rightarrow$DR&$\Lambda$CDM&  DM$\rightarrow$DR&$\Lambda$CDM\\
\hline\hline

{$\log(10^{10} A_\mathrm{s})$} & $3.044^{+0.015}_{-0.017}   $& $3.044\pm 0.016            $& $3.043^{+0.013}_{-0.015}   $& $3.041\pm 0.014            $& $3.051\pm 0.015            $& $3.049^{+0.013}_{-0.015}   $\\

{$n_\mathrm{s}   $} & $0.9644\pm 0.0043          $& $0.9641\pm 0.0044          $& $0.9676\pm 0.0038          $& $0.9679\pm 0.0036          $& $0.9700\pm 0.0041          $& $0.9711\pm 0.0036          $\\

{$\Omega_\mathrm{b} h^2$} & $0.02233\pm 0.00015        $& $0.02234\pm 0.00015        $& $0.02245\pm 0.00014        $& $0.02248\pm 0.00013        $& $0.02253^{+0.00017}_{-0.00014}$& $0.02261\pm 0.00013        $\\

{$\Omega_\mathrm{c} h^2$} & $0.1197\pm 0.0015          $& $0.1201\pm 0.0014          $& $0.1177^{+0.0014}_{-0.00070}$& $0.11829\pm 0.00082        $& $0.11695\pm 0.00091        $& $0.11713\pm 0.00079        $\\

{$\tau_\mathrm{reio}$} & $0.0542\pm 0.0080          $& $0.0541\pm 0.0079          $& $0.0553^{+0.0066}_{-0.0075}$& $0.0545\pm 0.0072          $& $0.0594^{+0.0068}_{-0.0080}$& $0.0594^{+0.0068}_{-0.0078}$\\

{$100\theta_\mathrm{s}$} & $1.04185\pm 0.00029        $& $1.04185\pm 0.00030        $& $1.04195\pm 0.00028        $& $1.04198\pm 0.00028        $& $1.04207\pm 0.00029        $& $1.04212\pm 0.00028        $\\
{$\log_{10}\zeta $} & $< -2.23                   $&---& $< -2.09                   $& ---&$-2.16\pm 0.47             $&---\\
{$\log_{10}(a_t) $} & $< 0.0340                  $& ---&$< 0.190                   $& ---&$< -0.731                  $&---\\

{$\kappa_{dcdm}  $} & $< 2.50                    $&---& $< 2.46                    $& ---&$< 2.51                    $&---\\\hline\hline

$\zeta                     $ & $0.00686^{-0.00044}_{-0.0063}$&---& $0.0102^{-0.0014}_{-0.0098}$&---& $0.01270^{+0.00053}_{-0.012}$&---\\

$H_0                       $ & $67.47\pm 0.64             $& $67.29\pm 0.61             $& $68.20\pm 0.41             $& $68.11\pm 0.38             $& $68.93^{+0.38}_{-0.50}     $& $68.70\pm 0.36             $\\

$\Omega_\mathrm{m}         $ & $0.3135\pm 0.0087          $& $0.3162\pm 0.0085          $& $0.3028\pm 0.0053          $& $0.3049\pm 0.0049          $& $0.2950^{+0.0054}_{-0.0048}$& $0.2975\pm 0.0046          $\\

$\sigma_8 (\Omega_\mathrm{m}/0.3)^{0.5}$ & $0.829\pm 0.016            $& $0.833\pm 0.016            $& $0.8091\pm 0.0091          $& $0.8107\pm 0.0088          $& $0.7993\pm 0.0086          $& $0.8006\pm 0.0085          $\\

\hline
\end{tabular}
\caption{The 68\% confidence limits on the base $\Lambda$CDM and DM$\rightarrow$DR parameters, along with the derived parameters $H_0$, $\Omega_m$, and $S_8$, for the DM$\rightarrow$DR and $\Lambda$CDM analyses with a linear prior on $\log\zeta$; cf.~Table~\ref{tab:all_confidence_intervals} for the analogous table for our baseline linear-$\zeta$-prior analysis. }\label{tab:all_confidence_intervals_logzeta}
\end{table*}

 \subsection{Linear prior on $\log\kappa$}

\begin{figure*}
 \includegraphics[width=\textwidth]{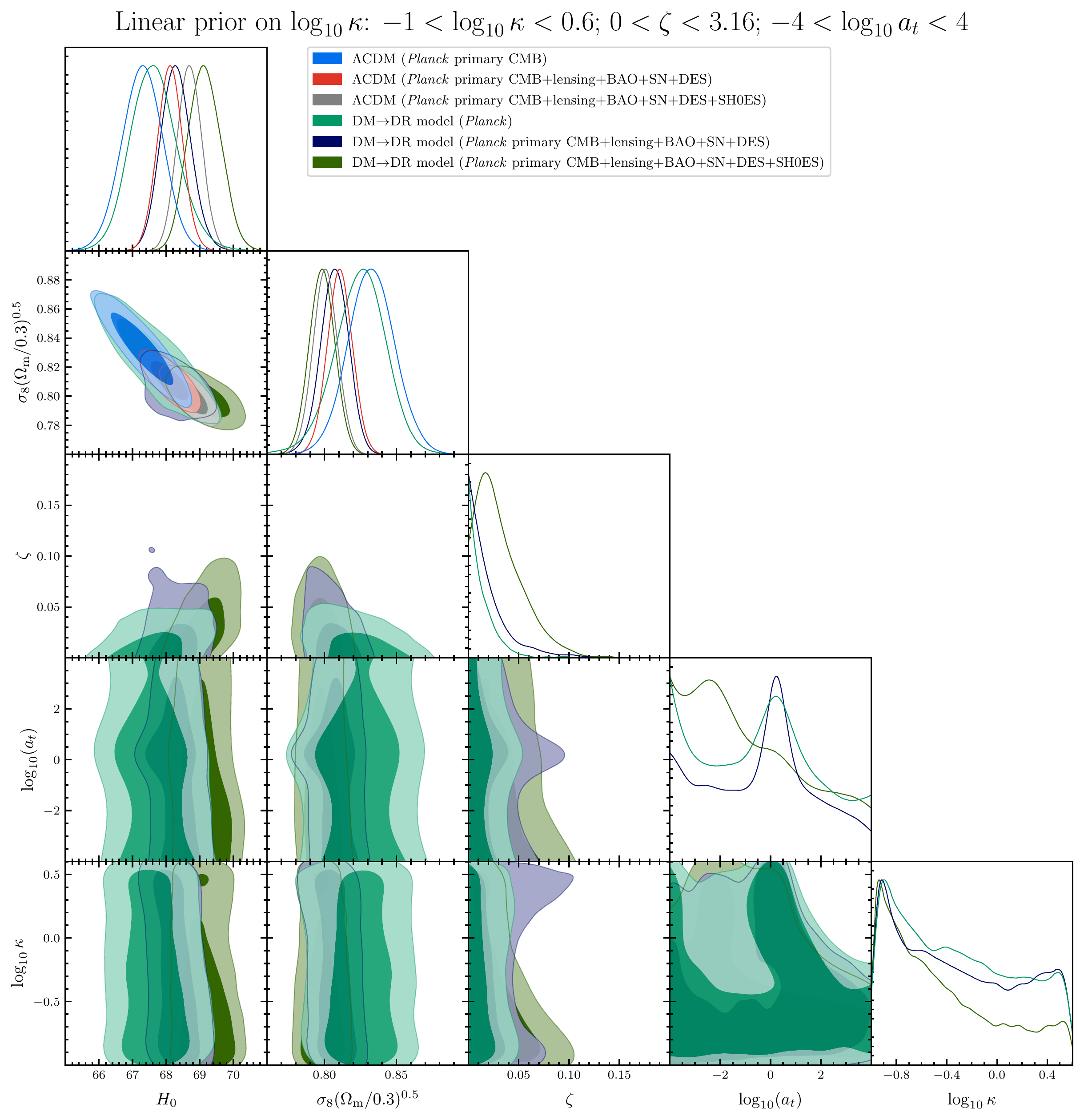}
 \caption{The posteriors on the DM$\rightarrow$DR parameters, when a linear prior on $\log_{10}\kappa$ is used (cf.~Figure~\ref{fig:posteriors}, where a linear prior on $\kappa$ is used).  }\label{fig:posteriors_logkappa}
 \end{figure*}

In Figure~\ref{fig:posteriors_logkappa} we present the posteriors when we use a linear prior on $\log_{10}\kappa$, while in the main text (and for our main results) we used a linear prior on $\kappa$. In particular, we impose $-1<\log_{10}\kappa<0.6$, which corresponds to our baseline linear prior of $0<\kappa<4$. We revert to a linear prior on $\zeta$, as in our main results: $0<\zeta<3.16$, and retain the linear prior on $\log a_t$: $-4<\log_{10}a_t<4$ and the physicality condition $\zeta<\frac{1}{a_t^\kappa}$.

 We present the 68\% confidence intervals on $H_0$ and $S_8$ in Table~\ref{tab:confidence_limits_logkappa} (in analogy to Table~\ref{tab:confidence_limits} for the baseline linear prior on $\kappa$). We present the 68\% confidence limits on the 6 base $\Lambda$CDM parameters and the DM$\rightarrow$DR parameters in Table~\ref{tab:all_confidence_intervals_logkappa}.
 
 While the general conclusions on $H_0$ and $S_8$ are unchanged, it is notable that the $\kappa$ and $\log_{10}a_t$ posteriors are slightly different in this case, as the sampler spends more time in the low-$\kappa$ regime which in turn explores the high-$a_t$ regime more than the linear-$\kappa$ case as a result of the physicality condition.

\begin{table*}
 \begin{tabular}{|c||c|c||c|c|}\hline
 & \multicolumn{2}{c||}{$H_0$ [km/s/Mpc]}&\multicolumn{2}{c|}{$S_8$}\\\cline{2-5}
 &$\Lambda$CDM& DM$\rightarrow$ DR &$\Lambda$CDM& DM$\rightarrow$ DR\\\hline\hline
\textit{Planck} primary CMB  &  $   66.69<H_0 <67.9   $  & $   66.9<H_0 <68.28   $  & $   0.82<S_8 <0.85   $  & $   0.81<S_8 <0.84   $  \\\hline 
+$\phi\phi$+BAO+SN+DES  &  $   67.74<H_0 <68.49   $  & $   67.82<H_0 <68.72   $  & $   0.80<S_8 <0.82   $  & $   0.80<S_8 <0.82   $  \\\hline 
+SH0ES  &  $   68.33<H_0 <69.06   $  & $   68.63<H_0 <69.61   $  & $   0.79<S_8 <0.81   $  & $   0.79<S_8 <0.81   $  \\\hline  
 \end{tabular}
 \caption{$68\%$ confidence limits on $H_0$ and $S_8$, from the analysis with a linear prior on $\log\kappa$; cf.~Table~\ref{tab:confidence_limits} for the analogous results for the baseline analysis with a linear prior on $\kappa$.}\label{tab:confidence_limits_logkappa}
 \end{table*}
  \begin{table*}
\begin{tabular} { |l|  c c |c c| c c|}\hline
&\multicolumn{2}{c|}{\textit{Planck} primary CMB}&\multicolumn{2}{c|}{$+\phi\phi+$BAO+SN+DES}&\multicolumn{2}{c|}{+SH0ES}\\\cline{2-7}
 Parameter &  DM$\rightarrow$DR&$\Lambda$CDM&  DM$\rightarrow$DR&$\Lambda$CDM&  DM$\rightarrow$DR&$\Lambda$CDM\\
\hline\hline

{$\log(10^{10} A_\mathrm{s})$} & $3.045^{+0.015}_{-0.017}   $& $3.044\pm 0.016            $& $3.045\pm 0.015            $& $3.041\pm 0.014            $& $3.055\pm 0.015            $& $3.049^{+0.013}_{-0.015}   $\\

{$n_\mathrm{s}   $} & $0.9643\pm 0.0045          $& $0.9641\pm 0.0044          $& $0.9668\pm 0.0039          $& $0.9679\pm 0.0036          $& $0.9687\pm 0.0039          $& $0.9711\pm 0.0036          $\\

{$\Omega_\mathrm{b} h^2$} & $0.02231\pm 0.00015        $& $0.02234\pm 0.00015        $& $0.02241\pm 0.00015        $& $0.02248\pm 0.00013        $& $0.02243\pm 0.00017        $& $0.02261\pm 0.00013        $\\

{$\Omega_\mathrm{c} h^2$} & $0.1190^{+0.0021}_{-0.0013}$& $0.1201\pm 0.0014          $& $0.1171^{+0.0020}_{-0.00047}$& $0.11829\pm 0.00082        $& $0.1166^{+0.0012}_{-0.00083}$& $0.11713\pm 0.00079        $\\

{$\tau_\mathrm{reio}$} & $0.0545\pm 0.0077          $& $0.0541\pm 0.0079          $& $0.0557\pm 0.0073          $& $0.0545\pm 0.0072          $& $0.0603^{+0.0070}_{-0.0080}$& $0.0594^{+0.0068}_{-0.0078}$\\

{$100\theta_\mathrm{s}$} & $1.04187\pm 0.00030        $& $1.04185\pm 0.00030        $& $1.04193\pm 0.00028        $& $1.04198\pm 0.00028        $& $1.04201\pm 0.00028        $& $1.04212\pm 0.00028        $\\
{$\log_{10}(\kappa_{dcdm})$} & $< -0.0628                 $&---& ---                         &---& $< -0.302                  $&---\\

{$\log_{10}(a_t) $} & $< 0.655                   $&---& $< 0.592                   $&---& $< 0.163                   $&---\\

{$\zeta          $} & $< 0.0166                  $&---& $< 0.0234                  $&---& $< 0.0394                  $&---\\\hline\hline

$\kappa                    $ & $0.091^{-0.066}_{-0.094}   $&---& $0.098^{-0.069}_{-0.10}    $&---& $0.131^{-0.035}_{-0.13}    $&---\\

$H_0                       $ & $67.65^{+0.63}_{-0.75}     $& $67.29\pm 0.61             $& $68.29\pm 0.45             $& $68.11\pm 0.38             $& $69.15\pm 0.49             $& $68.70\pm 0.36             $\\

$\Omega_\mathrm{m}         $ & $0.310^{+0.010}_{-0.0088}  $& $0.3162\pm 0.0085          $& $0.3006^{+0.0066}_{-0.0053}$& $0.3049\pm 0.0049          $& $0.2923\pm 0.0055          $& $0.2975\pm 0.0046          $\\

$\sigma_8 (\Omega_\mathrm{m}/0.3)^{0.5}$ & $0.825^{+0.019}_{-0.016}   $& $0.833\pm 0.016            $& $0.8070\pm 0.0098          $& $0.8107\pm 0.0088          $& $0.7983\pm 0.0088          $& $0.8006\pm 0.0085          $\\

\hline
\end{tabular}
\caption{The 68\% confidence limits on the base $\Lambda$CDM and DM$\rightarrow$DR parameters, along with the derived parameters $H_0$, $\Omega_m$, and $S_8$, for the DM$\rightarrow$DR and $\Lambda$CDM analyses with a linear prior on $\log\kappa$; cf.~Table~\ref{tab:all_confidence_intervals} for the analogous table for our baseline linear-$\kappa$-prior analysis. }\label{tab:all_confidence_intervals_logkappa}
\end{table*}

\subsection*{General conclusions}

While there is a slight shift in the 1-D posterior on $H_0$ between the linear--$\zeta$ and logarithmic-$\zeta$ prior cases, we find that the choice of prior does not affect our general conclusion that the DMDR model does not resolve cosmological tensions. We choose the linear priors on $\zeta$ and $\kappa$ for our main results in an effort to avoid the stretching of parameter space caused by the logarithmic priors at small values of the parameters; in particular, as both $\zeta \rightarrow 0$ and $\kappa \rightarrow 0$ are degenerate with $\Lambda$CDM, we wish to avoid the sampler spending inordinate time in a $\Lambda$CDM region by artificially stretching the prior volume there.

\bibliography{references}

\end{document}